\documentclass[preprint,11pt,authoryear]{elsarticle}

\newcommand{\rev}[1]{#1}

\newcommand{\appendixref}{\ref}

\usepackage{amsmath,amsthm,amsfonts,amssymb}
\usepackage{graphicx}
\usepackage{subcaption}
\usepackage[dvipsnames]{xcolor}

\usepackage{hyperref}
\hypersetup{
    colorlinks=true,
    linkcolor=blue,
	citecolor = blue,
    filecolor=magenta,      
    urlcolor=cyan
}

\usepackage[english]{babel}
\usepackage{float}

\newcommand{\bm}{\boldsymbol }

\newcommand{\pp}{\partial}
\newcommand{\dd}{\mr d}

\newcommand{\br}[1]{{ \left(  #1 \right) }}
\newcommand{\sbr}[1]{\left[ #1 \right]}

\newcommand{\lr}[3]{\left#1 #2 \right#3}

\let\underscore\_
\renewcommand{\_}[1]{_\mathrm{#1} }

\let\texthat\^
\renewcommand{\^}[1]{^\mathrm{#1} }

\newcommand{\mr}{\mathrm}
\newcommand{\mc}{\mathcal}

\newcommand{\ii}{\mr{i}}
\newcommand{\ee}{\mr{e}}

\DeclareMathOperator{\sech}{sech}

\newcommand{\bk}{{\bm k}}
\newcommand{\be}{{\bm e}}
\newcommand{\bx}{{\bm x}}
\newcommand{\bX}{{\bm X}}
\newcommand{\bu}{\bm u}
\newcommand{\h}{\hat}
\newcommand{\hX}{\hat X}
\newcommand{\w}{\omega}
\newcommand{\wn}{\w_n}
\newcommand{\wm}{\w_m}
\newcommand{\li}{\ell_i}

\newcommand{\oo}[1]{^{(#1)}}

\newcommand{\q}{\iota}

\newcommand{\ijn}{_{ijn}}
\newcommand{\jn}{_{jn}}
\newcommand{\kjn}{k\jn}
\newcommand{\kxjn}{k_{x,jn}}
\newcommand{\kyn}{k_{y,n}}
\newcommand{\bkjn}{\bm k_{jn}}
\newcommand{\lm}{_{lm}}
\newcommand{\klm}{k\lm}
\newcommand{\kxlm}{k_{x,lm}}

\newcommand{\bklm}{\bm k_{lm}}
\newcommand{\Kp}{K_{pnm}}
\newcommand{\Kxp}{K_{x,pnm}}

\newcommand{\bKp}{\bm K_{pnm}}

\newcommand{\cc}{c}

\newcommand{\symbsum}{+}
\newcommand{\symbdot}{\times}
\newcommand{\kdk}{k_{\symbdot}}
\newcommand{\akk}{k_{\symbsum}}
\newcommand{\bkk}{\bm k_{\symbsum}}
\newcommand{\kkx}{k_{x,\symbsum}}
\newcommand{\kky}{k_{y,\symbsum}}
\newcommand{\ww}{\w_{\symbsum}}
\newcommand{\inm}{_{inm}}
\newcommand{\jnlm}{_{jnlm}}

\newcommand{\pnm}{_{pnm}}
\newcommand{\qlm}{_{\q lm}}

\newcommand{\kap}{k}
\newcommand{\kkap}{\kappa}
\newcommand{\kapn}{\kappa_n}

\newcommand{\mdeg}{\text{\textdegree}}

\newcommand{\ain}{\h\eta_{i0n}\oo1}
\newcommand{\ai}{\h\eta_{i01}\oo1}
\newcommand{\A}{a}
\newcommand{\An}{a_n}
\newcommand{\wbl}{\delta} 

\newcommand{\td}{d}
\newcommand{\di}{\td_i}
\newcommand{\Nw}{N_\omega}

\journal{Coastal Engineering}

\begin{document}

\begin{frontmatter}
	
	\title{Second-order  theory for multi-hinged directional wavemakers} 
	\author{Andreas H. Akselsen}
	\affiliation{organization={SINTEF Ocean, Department of Ship and Ocean Structures},
		addressline={Paul Fjermstads vei 59}, 
		city={Trondheim},
		postcode={7052}, 
		state={Tr{\o}nderlag},
		country={Norway}}

\begin{abstract}
	\rev{
	The second-order directional wavemaker theory for regular and irregular waves is extended
	to multi-hinged wavemakers and combined piston--flap wavemaker systems.
	Derived expressions enable  second-order  signal correction, common in single-hinged wavemakers, to be applied to multi-hinged systems.
	Multi-hinged wavemakers offer additional degrees of freedom, with different combinations of paddle motion producing the same progressive wave.
	This is here exploited to better understand wavemaker behaviour.
	Single-harmonic  signals are computed for double-hinged wavemakers  that suppress spurious waves without introducing double-harmonic motions. 
	Surprisingly, these flap motions are almost always in opposite phase, with the larger draft found underneath the water surface. 
	Due to the opposing paddle phase, the double-hinged wavemaker draft is usually  smaller than the corresponding single-hinged draft.
	The ability of thsee systems  to suppress spurious waves with single-harmonic motion is  verified experimentally.
	The wavemaker theory further supports an  arbitrary number of flap  hinges, enabling the approximation of a fully flexible wavemaker through piecewise-linear segments.
	This is demonstrated with an exponential wavemaker profile that does not generate any evanescent waves at linear order. 
	Such a wavemaker is likely to limit wave breaking and cross-modes, but is found
	to produce spurious second-order waves of a magnitude comparable to a single flap.
	The presented solution is complete, intrinsically including return flow through the second-order zero mode. 
	This return flow is found to precisely match the Stokes drift.
}
\end{abstract}


\end{frontmatter}

\section{Introduction}

Wavemaker theory had its birth in the early 20'th century with \citet{havelock1929} and the linear transfer functions provided by \citet{biesel1951appareils} for flaps and pistons.
These functions predict accurate amplitudes when the wave steepness is small and the water is non-shallow. Accuracy diminishes with steeper waves, larger wavemaker draughts and shallower water depths.
Inaccuracies result in the formation of additional spurious, freely propagating waves which cannot be removed through calibration.

Weakly nonlinear wavemaker theories have been pursued in order to alleviate the problem of spurious (\textit{parasitic}) waves \citep{fontanet1961,madsen1971,sand1982,sand1985,barthel1983group}.
The most influential contribution came with \citet{schaffer_1996} who provided a theory that encompassed both regular and irregular waves for flap and piston wavemakers, including  transfer functions that suppress second-order spurious waves.
\citet{schaffer2003_3D_correction} later generalised the theory to directional wavefields as generated with multiple wavemaker flap aligned in the transverse direction.
Second-order wavemaker theory has since been examined  by \citet{pezzutto2016discontinuousWavemaker} to solve the convergence issue of discontinuous wavemaker geometries (an issue first noted by \citet{hudspeth1991stokes}). \citet{pezzutto2016discontinuousWavemaker} also provides the long desired formal proof of solution convergence. 

Second-order wavemaker theory, as based on the conventional Stokes expansion of variables, function well at deep-to-intermediate water depths. 
Weakly nonlinear theory is less suited to the shallow water range where the cascade of Stokes wave harmonics decays more slowly.
Approximate stream function wavemaker theory \citep{zhang2007streamFunctionWavemaker}, utilizing numerical Stokes wave solvers, provide  better control signals in the shallow-water range. 
\rev{A Boussinesq model has similarly been adopted for generating irregular  shallow-water  waves \citep{zhang2007_shallowPistonBoussinesq}.}
Both theories are limited to piston-type wavemakers. 
A review on the subject is provided by \citet{eldrup2019applicabilityWavemakerTheory}.
\\

\begin{figure}[H]%
\centering
\subfloat[Sketch of the double-hinged wavemaker used in SINTEF's Ocean Basin]{\includegraphics[width=.5\columnwidth]{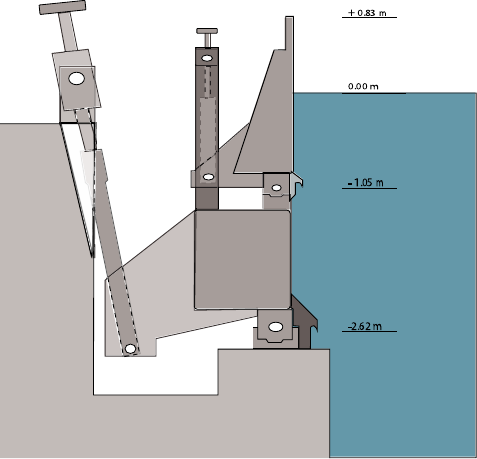}\label{fig:BM2}}%
\subfloat[Sketch of the multi-hinged wavemaker construct, annotated with sybols used in \ref{sec:BC_x}. Angles are exaggerated.]{%
\parbox{.5\columnwidth}{\centering\includegraphics[width=.25\columnwidth]{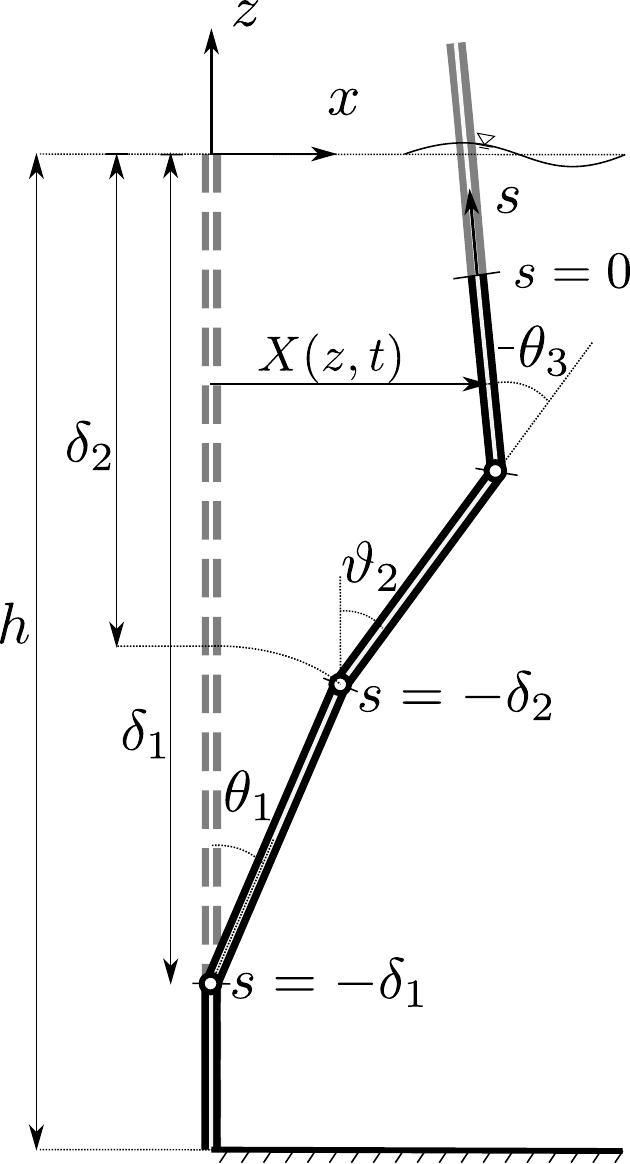}}\label{fig:flap}}%
\caption{Multi-hinge wavemaker.}%
\label{fig:multiflap}%
\end{figure}

\rev{This paper provides second-order wavemaker theory for a category of wavemakers that is not well studied in literature.   
	Although less common than the piston or the single-hinged flap, the double-hinged flap wavemaker can be found in hydrodynamic laboratories around the world today, for example at SINTEF Ocean (\autoref{fig:BM2},\ref{fig:BM2photo}) and in the Indonesian Hydrodynamic Laboratory (IHL) \citep{kusumawinahyu_2017_double_flap}.
	To generalise, an arbitrary number of hinges are assumed, the infinitely deep hinge functioning as a piston,
	and the collective noun `paddle' is adopted for both piston and flap.
	Ever-increasing wavemaker flexibility may be attained by increasing the number of wavemaker hinges, thus providing a piecewise-linear approximation of  smooth paddle shape.
   The derived theory, which  follows in the foots of \citet{schaffer2003_3D_correction},  supports multi-directional wave fields as may be generated by multiple paddles aligned in the spanwise direction.
}

This paper is organised as follows:
Theory is derived and validated in \autoref{sec:theory} and then utilised in \autoref{sec:optimize} to find the  flap motion that eliminates regular spurious waves  without imposing higher frequencies. 
The theory is validated experimentally in \autoref{sec:experiment} using this special single-harmonic motion.
Fully flexible wavemakers are considered in \autoref{sec:flexi} and a summary follows in \autoref{sec:summary}.

\section{Second-order wavemaker model}
\label{sec:theory}

\subsection{Governing equations}

In a coordinate system oriented along the still water level and upright wavemaker, the Laplace equation for potential flow, accompanied with boundary conditions, reads
\begin{subequations}
\begin{alignat}{2}
\nabla^2 \phi &= 0,&&
\\
\phi_t + \tfrac12|\bu|^2 + g\eta &= 0 &\quad \text{at } z&=\eta,\label{eq:BC:dyn}
\\
\phi_{tt} + g \phi_z + \big(|\bu|^2\big)_t + \tfrac12\bu\cdot\nabla\big(|\bu|^2\big)  &= 0 &\quad \text{at } z&=\eta,\label{eq:BC:dynCombo}
\\
\frac{D}{Dt}(x-X) &= 0 & \text{at } x &= X(y,x,t), \label{eq:BC:X}
\\ 
\phi_z &= 0 &\quad \text{at } z&=-h,
\\
\phi, \eta &\text{ finite}& \text{as }x&\to\infty,\, y\to\pm\infty,
\end{alignat}%
\label{eq:system}%
\end{subequations}%
where
$\bu = \nabla \phi$.
The kinematic boundary condition $\frac{D}{Dt}(z-\eta)=0$ at $z=\eta$ has here been combined with the dynamic condition \eqref{eq:BC:dyn} to eliminate $\eta$, yielding \eqref{eq:BC:dynCombo}.
The Stokes expansion together with a Taylor expansion about the respective reference planes $x=0$ and $z=0$ generates a cascade of repeating linear systems to be solved at each order of approximation. 
Following \citet{schaffer_1996}, we write the resulting linearised equations on the form
\begin{subequations}
\begin{alignat}{2}
\nabla^2 \phi\oo m &= 0,&&\label{eq:system_lin:Laplace}\\
\phi_t\oo m+  g \eta\oo m   &= P\oo m &\quad \text{at } z&=0,  \label{eq:system_lin:P}\\
\phi_{tt}\oo m + g \phi_z\oo m  &= R\oo m &\quad \text{at } z&=0,\label{eq:system_lin:R}\\
\phi_x\oo m &= X_t\oo m + Q\oo m& \text{at } x&=0,\label{eq:system_lin:Q}\\
\phi_x\oo m &= 0 &\quad \text{at } z&=-h,\label{eq:system_lin:z=h}\\
\phi\oo m, \eta\oo m &\text{ finite}& \text{as }x&\to\infty,\, y\to\pm\infty\label{eq:system_lin:x=inf}
\end{alignat}%
\label{eq:system_lin}%
\end{subequations}
for approximation orders $m = 1,2,\ldots$
The right-hand terms are combinations of known lower-order functions. To second order, we have
\begin{subequations}
\begin{align}
P\oo1 &= 0, & P\oo2 &= -\eta\oo 1 \phi_{zt}\oo1 - \tfrac12 \big|\bu\oo1\big|^2, &\text{at } z&=0,\label{eq:defP}\\
R\oo1 &= 0, & R\oo2 &= -\big(\big|\bu\oo1\big|^2\big)_t - \eta\oo1 (\pp_{tt}+g\pp_z)\phi\oo1_z & \text{at } z&=0,\label{eq:defR}\\
Q\oo1 &= 0, & Q\oo2 &= \bu\oo1\cdot \nabla X\oo1 - X\oo1 \phi_{xx}\oo1 &\text{at } x&=0.\label{eq:defQ}
\end{align}%
\label{eq:defPRQ}%
\end{subequations}%

\subsection{The first-order solution}
\label{sec:O1}

The multi-hinge wavemaker construct is sketched in \autoref{fig:flap}.
\citet{kusumawinahyu_2017_double_flap} observed for linear theory that the multi-hinge boundary condition is a superposition of mono-hinge conditions.
We show in \ref{sec:BC_x} that this observation  holds up to second-order theory but not above.
Similarly, the linear solution becomes a superposition of the mono-hinge solutions respective to each hinge, with
\begin{subequations}
\begin{align}
X\oo1(y,z,t) &= \frac12 \sum_{i=1}^N  \sum_n \hX_{in}\oo1 \frac{\max(\wbl_i+z,0)}{\wbl_i} \ee^{\ii (\wn t - \kyn y)},\label{eq:XO1}\\
\phi\oo1(x,y,z,t) &= \frac12 \sum_{i=1}^N \sum_n \sum_{j=0}^{\infty} \h\phi_{ijn}\oo1 \frac{\cosh \kjn (z+h)}{\cosh \kjn h}
\ee^{\ii(\wn t - \bkjn \cdot\bx)},\label{eq:phiO1}\\ 
\eta\oo1(x,y,t) &= \frac12\sum_{i=1}^N  \sum_n \sum_{j=0}^{\infty} \h\eta_{ijn}\oo1\ee^{\ii(\wn t - \bkjn \cdot\bx)}, \label{eq:etaO1}
\end{align}%
\label{eq:phieta_O1}%
\end{subequations}%
and  $\bkjn = (\kxjn,\kyn,0)$. 
The depth down to hinge $i$ is denoted $\wbl_i$ (positive). 
Paddle draught $\hX_{in}\oo1$ or, equivalently, the linear progressive wave amplitudes $\h\eta_{i0n}\oo1$ are normally the prescribed input parameters.
\\

The first-order surface boundary condition \eqref{eq:system_lin:R} provides the relationship between wavelengths and wave frequencies, and is the well-known dispersion relation 
\begin{equation}
\wn^2 = g \kjn \tanh \kjn h.
\label{eq:dispersion}
\end{equation} 
Wavenumber moduli $\kjn $ of the primary free waves are obtained as solutions to the dispersion relation given wavemaker frequency $\wn$.
Two real and infinitely many imaginary wavenumbers exist satisfying  \eqref{eq:dispersion}. 
Real wavenumbers are associated with progressive waves while the imaginary ones forms the evanescent near field surrounding the wavemaker. 
By common conversion, we let negative frequency indices $n<0$ correspond to negative frequencies 
and arrange
$\kjn $ such that $j=0$ is the real solution of the dispersion relation 
while $j>0$ are the imaginary ones.

The spanwise wavenumber component $\kyn$ is a real input parameter from which the longitudinal wavenumber
\begin{equation}
\kxjn=\pm \sqrt{\kjn^2-\kyn^2}
\label{eq:kx}
\end{equation}
is prescribed.  
The negative imaginary branch  $\ii \,\kxjn >0$ must be chosen whenever $\kxjn$ is imaginary.
When $\kxjn$ is real, the branch matching the sign of the frequency must be chosen.  
Note that all longitudinal wavenumbers become imaginary if $|\kyn|>|k_{0n}|$, leaving only evanescent modes.
If progressive waves are present, then their direction is 
\begin{equation}
	\alpha_n = \arcsin(\kyn/k_{0n}).
	\label{eq:alpha}
\end{equation}
Indices related to directionality are here absorbed in the frequency index $n$.
Components of negative frequency $n<0$ are always complex conjugates of oners. 

The remaining first-order amplitudes are obtained from \eqref{eq:system_lin:P},
directly giving
\begin{equation}
\h\eta\ijn\oo1 = - \ii  \frac{\wn}{g} \h\phi\ijn\oo1.
\end{equation}
Finally, these amplitudes are related to the flap motion via \eqref{eq:system_lin:Q}.
Writing 
\begin{equation}
\h\eta\ijn\oo1 = \ii \hX_{in}\oo1 \cc_i(\bkjn), 
\label{eq:heta_O1}
\end{equation}
the Bi{\'e}sel transfer function $\cc_i$ is found by multiplying \eqref{eq:system_lin:Q} with its orthogonal basis, which is
 $\cosh k_{ln}(h+z)\sech k_{ln}h$,
 and integrating from $z=-h$ to $0$.
The result is
\begin{align}
\cc_i(\bm k) = \frac{\tanh k h}{\Lambda(\bm k)}\Gamma_{i,1}(k),
\label{eq:c}
\end{align}
kernel $\Lambda$ and shape functions $\Gamma_{i,1}$  given in \appendixref{sec:shape_functions}.

\subsection{The second-order solution}
\label{sec:O2}
\begin{figure}%
\centering
\includegraphics[width=.4\columnwidth]{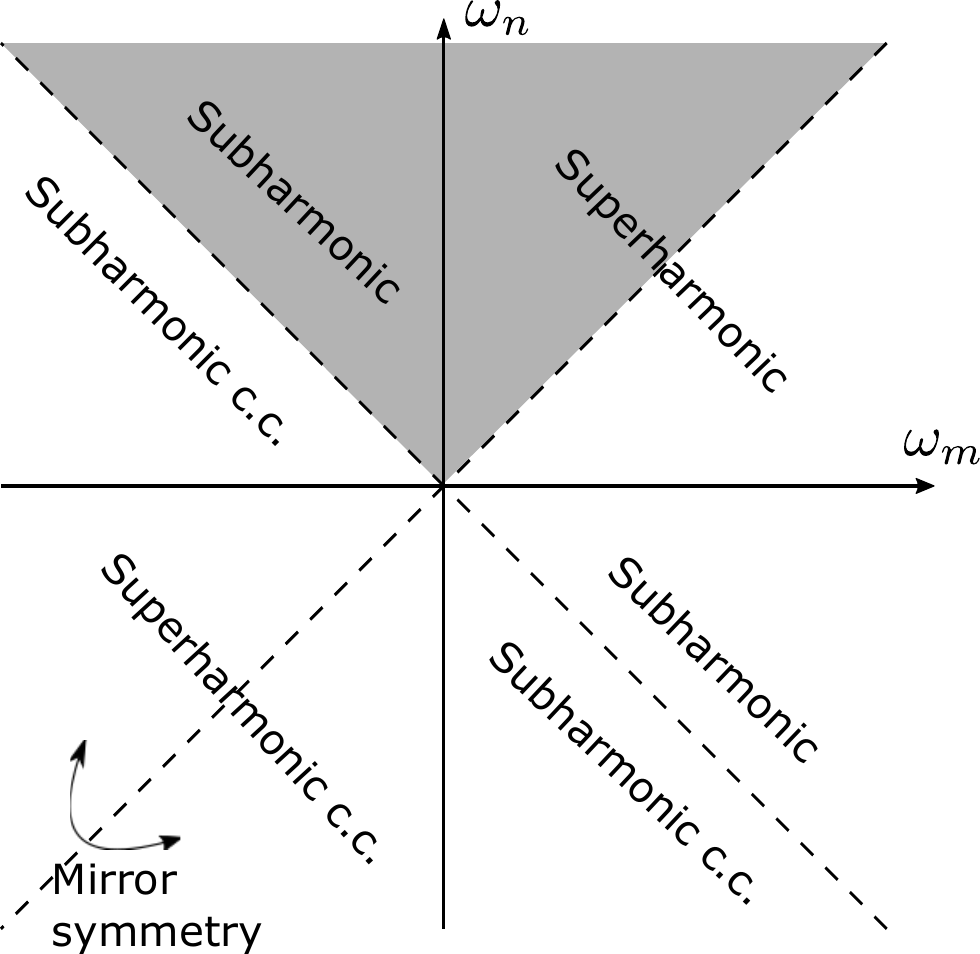}%
\caption{Space of interaction frequencies. Shaded area indicate the minimal domain of computation.}%
\label{fig:omega_space}%
\end{figure}

The second-order solution is comprised of interactions between first-order modes. 
These will be of frequency $\wn  + \wm$, 
and an index triplet $\q lm$ (hinge, mode, frequency) corresponding to  $ijn$  is introduced for interacting cross terms.
Superharmonics constitute interactions between frequencies of the same sign and subharmonics interactions of opposite sign.
\citet{schaffer_1996} adopts a $\pm$-suffix to distinguish between sub- and superharmonics. 
We will instead simply express the system over the whole domain of interactions, noting the domain symmetries as illustrated in \autoref{fig:omega_space}.
Apart from the diagonals,\footnote{$|n|=|m|$ which Sch{\"a}ffer's accounted for with a tensor `$\delta_{nm}$'.} the minimal domain of computation, which is shaded gray in \autoref{fig:omega_space}, is  one-quarter of the full domain.
Similar mirror symmetry exists with respect to hinge interactions  $i$ and $\q$.

Following Sch{\"a}ffer's notation, we
split the second-order solution into its bound harmonics (annotated 21), its (parasitic)  free harmonics (annotated 22), and any additional paddle motion  introduced  to suppress the latter (annotated 23):
\[
\phi\oo2 =\phi\oo{21}+\phi\oo{22}+\phi\oo{23}
;\qquad
\eta\oo2 =\eta\oo{21}+\eta\oo{22}+\eta\oo{23}
\]
These account for separate parts of the boundary value problem \eqref{eq:system_lin}, solving the sub-problems
\begin{equation}
\left\{\begin{alignedat}{2}
\phi_t\oo{21}+  g \eta\oo{21}  &= P\oo 2 &\quad \text{at } z&=0\\
\phi_{tt}\oo{21} + g \phi_z\oo{21} &= R\oo2 &\quad \text{at } z&=0\\
&\text{arbitrary}& \text{at } x&=0
\end{alignedat}%
\right\},
\label{eq:system_lin:21}
\end{equation}
\begin{equation}
\left\{\begin{alignedat}{2}
\phi_t\oo{22}+  g \eta\oo{22}  &=0 &\quad \text{at } z&=0\\
\phi_{tt}\oo{22} + g \phi_z\oo{22} &=0 &\quad \text{at } z&=0\\
\phi_x\oo{22} &= -\phi_x\oo{21} + Q\oo2& \text{at } x&=0
\end{alignedat}%
\right\},
\label{eq:system_lin:22}
\end{equation}
\begin{equation}
\left\{\begin{alignedat}{2}
\phi_t\oo{23}+  g \eta\oo{23}  &= 0 &\quad \text{at } z&=0\\
\phi_{tt}\oo{23} + g \phi_z\oo{3} &= 0 &\quad \text{at } z&=0\\
\phi_x\oo{23} &= X_t\oo{2}& \text{at } x&=0
\end{alignedat}\right\}%
\label{eq:system_lin:23}
\end{equation}
alongside \eqref{eq:system_lin:Laplace}, \eqref{eq:system_lin:z=h} and \eqref{eq:system_lin:x=inf}.

\subsubsection*{The second-order bound wave (21)}
\label{sec:O21}
The solution to the bound wave 
problem \eqref{eq:system_lin:21} is 
\begin{subequations}
\begin{align}
\phi\oo{21} &= \sum\jnlm\frac12 \h\phi\jnlm\oo{21}\frac{\cosh \akk (z+h)}{\cosh \akk h}
\ee^{\ii(\ww t-\bkk \bx )},\label{eq:phi21}%
\\
\eta\oo{21} &= \sum_{jnlm}\frac12 \h\eta\jnlm\oo{21}\ee^{\ii(\ww t-\bkk \bx )},\label{eq:eta21}%
\end{align}%
\label{eq:phietaO21}%
\end{subequations}%
where the shorthands
\begin{align*}
&&\ww &= \wn+\wm, &&\\
\bkk &= \bk\jn + \bk\lm, &
\akk &= (\bkk\cdot\bkk)^{1/2}, &
\kdk &= (\bk\jn\cdot\bk\lm)^{1/2}
\end{align*}
have been introduced  and sum run over all prescribed indices. 
The amplitudes follow directly form \eqref{eq:system_lin:21}: 
\begin{align}
\h\phi\jnlm\oo{21} &= \frac{ \h R\jnlm\oo2}{g \akk \tanh \akk h - \ww^2},&
\h\eta\jnlm\oo{21} &=  \frac{1}{g} \br{-\ii \ww \h\phi\jnlm\oo{21} + \h P\jnlm\oo2}.
\label{eq:hphiO21}
\end{align}%
Evaluating the source terms is a little more labour-intensive; one finds
\begin{subequations}
\begin{align}
\h R\jnlm\oo2 &= \ii\sum_{i,\ii=0}^N\h\eta\ijn\oo1\h\eta\qlm\oo1\sbr{
 \ww\br{\wn \wm - \frac{g \kdk^2}{\wn\wm}}+ \frac{\wn^3+\wm^3}{2}-\frac{g^2}{2}\br{\frac{\kjn^2}{\wn } + \frac{\klm^2}{\wm }}},\label{eq:RO2} \\
\h P\jnlm\oo2 &= -\sum_{i,\ii=0}^N\h\eta\ijn\oo1\h\eta\qlm\oo1\sbr{ \frac12 \br{ \frac{g^2 \kdk^2}{\wn\wm} + \wn\wm - \ww^2 }}.\label{eq:PO2}
\end{align}%
\end{subequations}%
Square bracket terms correspond to Sch{\"a}ffer's `$H\jnlm$'  and `$L\jnlm$' tensors. 
\subsubsection*{The spurious free wave (22)}
The free waves of the problem \eqref{eq:system_lin:22} take the form
\begin{subequations}
\begin{align}
\phi\oo{22} &= \frac12 \sum\pnm\h\phi\pnm\oo{22} \frac{\cosh \Kp (z+h)}{\cosh \Kp  h}  \ee^{\ii(\ww t-\bKp \cdot\bx)}, \label{eq:phiO22}%
\\
\eta\oo{22} &= \frac12 \sum\pnm  \h\eta\pnm\oo{22}  \ee^{\ii(\ww t-\bKp \cdot\bx)}; \quad \h\eta\pnm\oo{22}= -\ii \frac{\ww}{g} \h\phi\pnm\oo{22} \label{eq:etaO22}%
\end{align}%
\label{eq:phietaO22}%
\end{subequations}%
with
$\bKp = (\Kxp,\kky,0)$,
$\Kp $ 
 being the wavenumber moduli spurious waves. 
These are freely dispersing linear waves formed as a result of second-order inconsistencies at the wavemaker---they obey
\begin{equation}
\ww^2 = g \Kp \tanh \Kp h.
\label{eq:dispersionK}
\end{equation} 
As with the first-order wave, $p=0$ corresponds to the positive real root while $p=1,2,\ldots$ to the imaginary roots, 
and $\Kxp=\pm[\Kp^2-\kky^2]^{1/2}$ with branches chosen as with $\kxjn$.

Similar to earlier, 
the orthogonality of \eqref{eq:orthCosh} in combination with \eqref{eq:dispersionK} now yields%
\begin{align}
&\h\phi\pnm\oo{22} = \frac{1}{\Lambda(\bKp)}
\bigg[
\frac{\ii}{4}\sum_{i,\q=1}^N\sum_{j=0}^\infty \h X_{\q m}\oo1 \h\phi\ijn\oo1 \kjn\Gamma_{\q,2}(\bkjn,\bKp)
\nonumber\\&\phantom{=}\
+\frac{\ii}{4}\sum_{i,\q=1}^N\sum_{l=0}^\infty \h X_{in}\oo1 \h\phi\qlm\oo1  \klm\Gamma_{i,2}(\bklm,\bKp)
-\frac{1}{g} \sum_{j,l=0}^\infty   \frac{\kkx}{\akk^2-\Kp^2}\h R\jnlm
\bigg].
\label{eq:hphiO22}
\end{align}
Shape functions $\Gamma_{i,2}$ are given in \appendixref{sec:shape_functions}---%
note the index $\q$ appearing with the argument $\bkjn$ and $i$ with $\bklm$ in \eqref{eq:hphiO22}.
As mantioned earlier, 
computation can be reduced by exploiting the mirror symmetries.

\subsubsection*{The wave correction (23)}
\label{sec:O23}
The correction wave is a solution to \eqref{eq:system_lin:23}, which is analogous to the first order problem. 
Correction is imposed as an addition to the paddle motion
$X = X\oo1+X\oo{2}$,
where
\begin{equation}
X\oo{2}(y,z,t) = \frac12  \sum_{nm} \sum_{i=1}^N \hX\inm\oo{2} \frac{\max(\wbl_i+z,0)}{\wbl_i} \ee^{\ii (\ww t-\kky y)}.
\label{eq:XO2}
\end{equation}
Analogous to \eqref{eq:c}, this motion generates waves of amplitude
\begin{equation}
\h\eta\pnm\oo{23} = \sum_{i=1}^N\ii \hX\inm\oo2 c_i(\bm K\pnm)
\end{equation}
with fields
\begin{subequations}
\begin{align}
 \phi\oo{23} &= \frac12 \sum_{nm} \sum_{p=0}^\infty  \h\phi\pnm\oo{23}  \frac{\cosh \Kp (z+h)}{\cosh \Kp  h}\ee^{\ii (\ww t- \bKp \cdot\bx)}, \label{eq:phiO23}%
\\
\eta\oo{23} &= \frac12 \sum_{nm}  \sum_{p=0}^\infty \h\eta_{pnm}\oo{23} \ee^{\ii (\ww t-\bKp \cdot\bx)}; \quad \h\eta\pnm\oo{23}= -\ii \frac{\ww}{g} \h\phi\pnm\oo{23}.\label{eq:etaO23}
\end{align}%
\label{eq:phietaO23}%
\end{subequations}%
Note that the correction itself also generates spurious waves at third order and above.

Cancelling the progressive component of $\eta\oo{22}$ with the progressive component of $\eta\oo{23}$, we get
\begin{equation}
\sum_{i=1}^N  \ii\, c_i(\bm K_{0nm}) \hX_{inm}\oo{2}= -\h\eta_{0nm}\oo{22}.
\label{eq:heta23_sum_i}
\end{equation}
The problem is underdetermined when more than one hinge is present, 
and one is free to choose how the  correction is weighted across the paddle segments. 
Introducing hinge weights $w_i$, condition \eqref{eq:heta23_sum_i} is explicitly written 
\begin{equation}
\hX\oo{23}_{inm}=  w_i\frac{\ii\h\eta_{0nm}\oo{22}}{c_i(\bm K_{0nm})};\quad \sum_{i=1}^N w_i = 1.0.
\label{eq:hX_weight}
\end{equation}
Chosen  weights should ideally minimize the spurious waves generated at higher orders. 
As we shall see later, out-of-phase motion is preferable, meaning $w_i\in\mathbb C$.

\subsection{Special limits}
\label{sec:limits}
Singularities arise in the second-order solution.
These have physical significance.
Roots in  \eqref{eq:hphiO21}, signifying coincidence of bound and free waves, have been extensively studied in literature.
At third order, they are encountered with  $\ww\neq0$, where these play a central role in wave stability and higher order wave dispersion. 
At second order, only roots for which $\ww\to0$ appear. These constitute  a set-down in water level which can be expressed
\begin{equation}
\lim_{\wm\to-\wn} \h\eta_{i0n\q 0m}\oo{21} = \h\eta_{i0n}\oo1\h\eta_{\q0n}\oo1
\frac{ 4 k_{0n}\cosh^2 \kapn\,(4\kapn + \sinh 2 \kapn)}{8 \kapn^2 -4 \kapn \sinh 4\kapn + \cosh 4 \kapn -1},
\end{equation}
 $\kapn = k_{0n}h$.
Set-down may be interpreted as the limit case for when the mean surface level decreases around a long wave packet. 
It is also associated with a uniform current
\begin{equation}
\lim_{\wm\to-\wn}\!\!\! - \ii (k_{0n}+k_{0m})\h\phi_{i0n\q0m}\oo{21} =\h\eta_{i0n}\oo1\h\eta_{\q0n}\oo1
\frac{2 g k_{0n}^2}{\w_n}
\frac{ 2 \kapn + 3\sinh 2 \kapn + \sinh 4\kapn}{8 \kapn^2 -4 \kapn \sinh 4\kapn + \cosh 4 \kapn -1},
\label{eq:U021}
\end{equation}
both of which are merely convergence limits of \eqref{eq:hphiO21} and are not required in a particular solution 
since \eqref{eq:system_lin:21} will be satisfied with any arbitrary zero mode. 
The limits can however be useful for generating smooth transfer functions.
\\

Though often neglected, a complete solution also requires careful consideration of roots in the spurious free modes \eqref{eq:hphiO22}.
According to the dispersion relation \eqref{eq:dispersionK},
free zero-harmonic wavenumber are 
\begin{equation}
K_{pn-n} =  \frac{\ii \pi p}{h},
\end{equation}
which are seen to be stationary Fourier harmonics with nodes at $z = 0$ and $z=-h$.
The corresponding progressive component $K_{0n-n} = 0$ is a root which cannot be computed directly from \eqref{eq:hphiO22},
yet it appears as part of  the orthogonal basis for $\h\phi\pnm\oo{22}$.
It is however the derivative $\phi_x\oo{22}$ which enters the  lateral boundary condition \eqref{eq:system_lin:Q} and not $\phi\oo{22}$ itself, 
and one may therefore include the zeroth harmonic by adding a uniform current $U_0 \bm e_x$  directly to the velocity field:
\begin{equation}
\bm u = \nabla \phi + U_0 \bm e_x.
\end{equation}
Setting $\h\phi_{0n-n}\oo{22} = 0$,
one defines 
\begin{equation}
U_0 =\frac12\sum_n \lim_{\wm\to-\wn} -\ii K_{x,0nm}\h\phi_{0nm}	\oo{22},
\end{equation}
which, observing that 
$
\lim_{K_x\to 0} K_x /\Lambda(\bm K) = 1/h$
and $\Gamma_{i,2}(\bk,0) = \tanh(k h)$,
evaluates to
\newcommand{\mn}{-n}
\begin{align}
 U_0 &
= \frac1{4h}\sum_n \sum_{i,\q=1}^N 
\bigg[
\sum_{j=0}^\infty \h X_{\q \mn}\oo1 \h\phi\ijn\oo1 \kjn\tanh(k_{jn} h) 
+\frac{g}{2\wn}\sum_{\substack{j,l=0\\j+l\neq0}}^\infty  \h\eta\ijn\oo1\h\eta\qlm\oo1 (\kxjn-\kxlm)
\bigg].
\label{eq:U0}
\end{align}
The $j=l=0$ component of the last term is here excluded because this component essentially functions to cancel any arbitrary current in the partial solution of \eqref{eq:system_lin:21};
it may be neglected provided we also negate  \eqref{eq:U021}.
\\

\begin{figure}[!hptb]%
\centering
\includegraphics[width=.60\columnwidth]{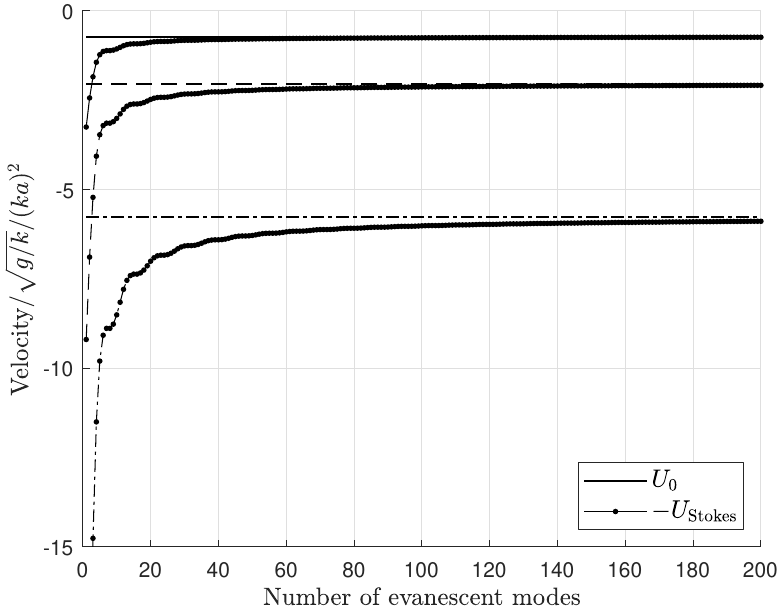}%
\caption{Convergence of back-flow velocity \eqref{eq:U0} to the Stokes drift velocity \eqref{eq:U_stokes} for 
double-hinged wavemaker $\wbl_i/h=[\frac14,\frac34]^T$, $k_{01}\ai/ka=[\frac23,\frac13]^T$.
Solid: $kh=\frac\pi4$, dashed: $kh=\frac\pi8$, dot-dashed: $kh=\frac\pi{16}$.
}%
\label{fig:U_of_nEv}%
\end{figure}

So what is the significance of $U_0$?
At first glance, the fixed current  may seem to violate the impermeability of the wavemaker, but is only from the Eulerian point of view. 
In the Lagrangian frame, the second-order solution provides fluid transport through Stokes drift.
Mass conservation therefore dictates that $U_0$ should equal the Stokes drift in magnitude and flow in the opposing direction. 
Indeed, to second order, the Stokes drift is
\begin{equation}
U_\mathrm{Stokes} 
=\frac{1}{h} \bigg\langle \int_0^\eta \!\phi_x\,\dd z\bigg\rangle
=\frac g{4h}\sum_n \frac{k_{x,0n}}{\wn}\sum_{i,\q=1}^N \h\eta_{i0n}\oo1\big(\h\eta_{\q0n}\oo1\big)^* + O\big(|\h\eta_{i0n}|^3\big),
\label{eq:U_stokes}
\end{equation}
and  \autoref{fig:U_of_nEv} demonstrates that  this is the value to which  \eqref{eq:U0} converges.
Similar convergence in   single-hinge wavemaker theory was found by \citet{hudspeth1991stokes}.

\subsection{Validation}
\label{sec:validation}

\begin{figure}[H]%
	\centering
	\subfloat[First order lateral boundary condition.]{\includegraphics[width=.5\columnwidth]{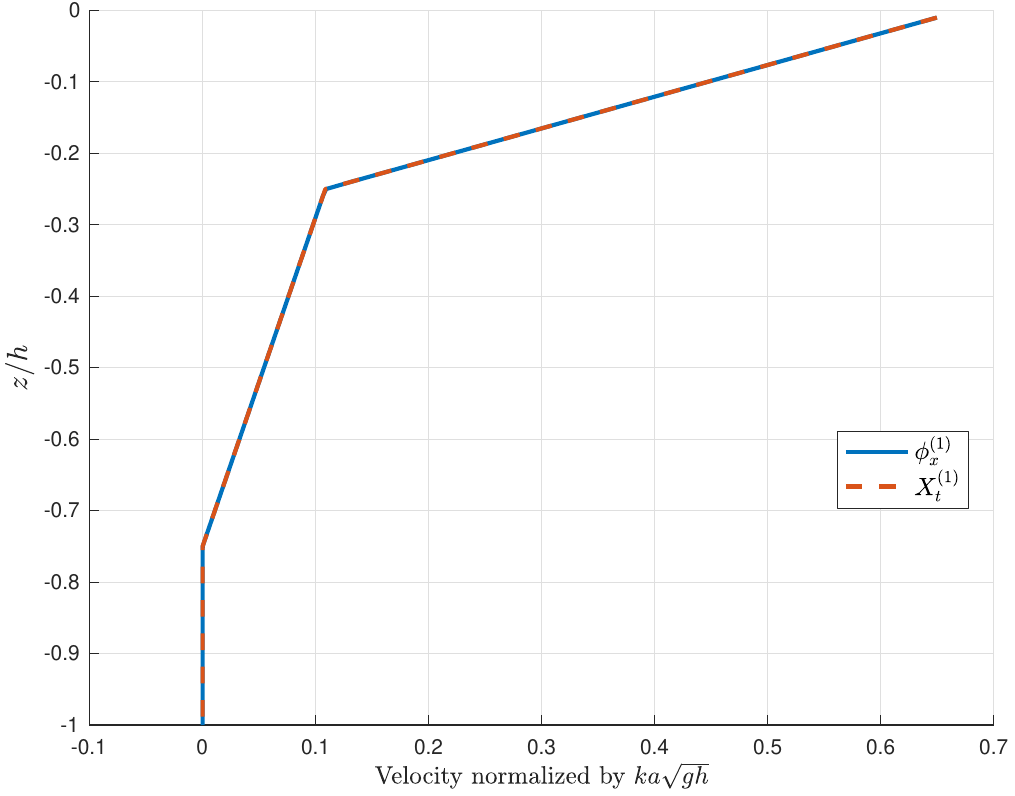}\label{fig:validation:w2wb2:QO1}}%
	\subfloat[second-order lateral boundary condition.]{\includegraphics[width=.5\columnwidth]{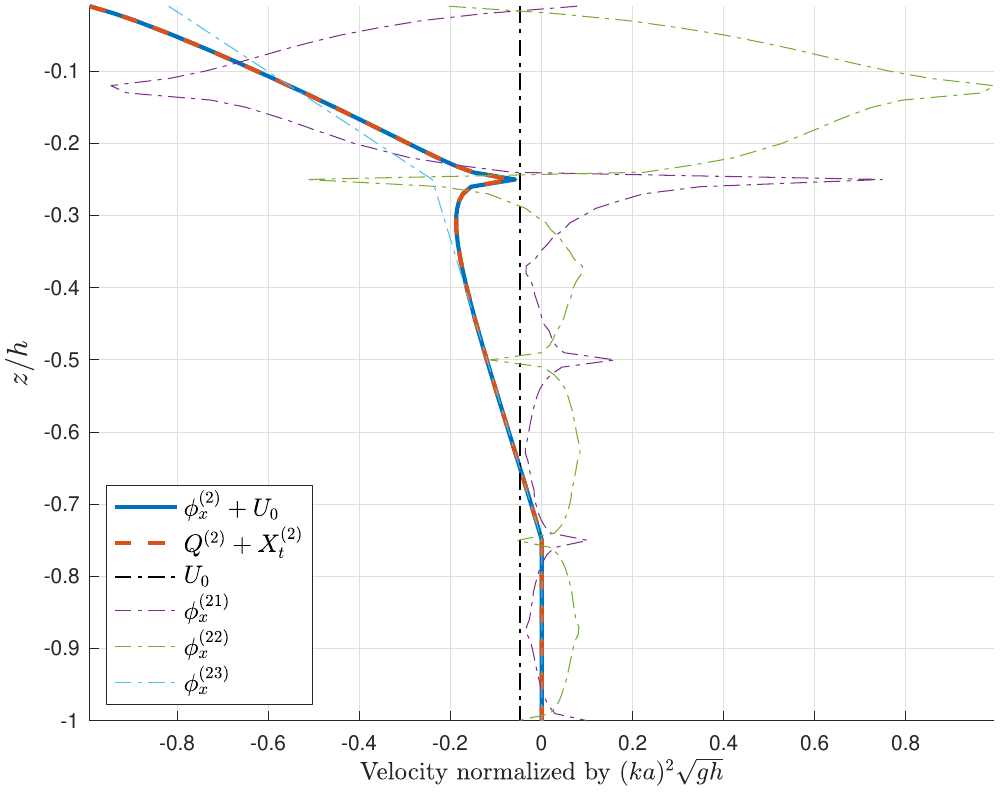}\label{fig:validation:w2wb2:QO2}}\\
	\subfloat[second-order kinematic condition.]{\includegraphics[width=.5\columnwidth]{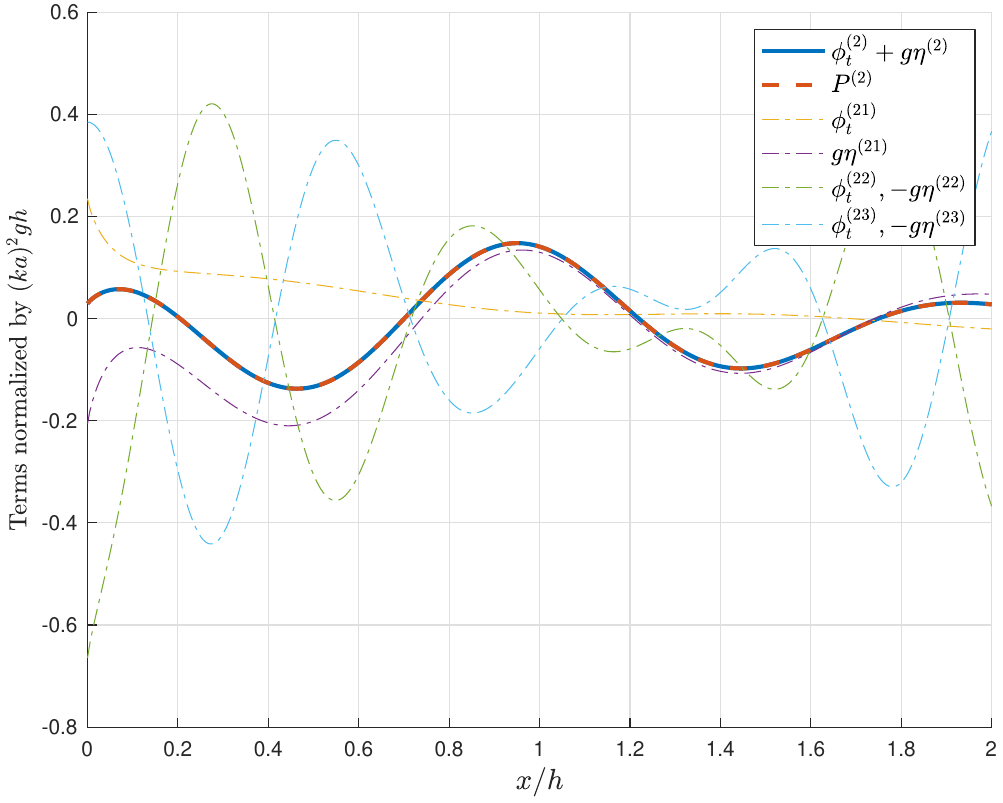}\label{fig:validation:w2wb2:PO2}}%
	\subfloat[second-order dynamic condition.]{\includegraphics[width=.5\columnwidth]{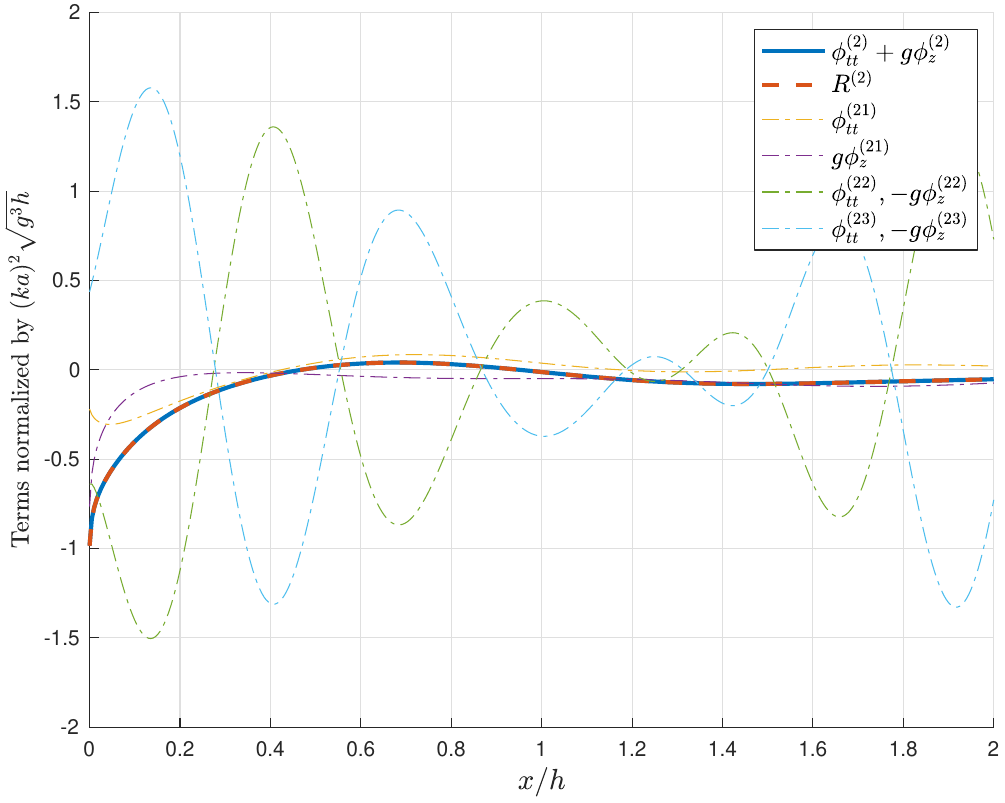}\label{fig:validation:w2wb2:RO2}}%
	\caption{Example validation for a double-hinged flap moving at different frequencies. For $n=1,2$, 
		$\wn\sqrt{h/g} = [2.0,1.5]$, $\wbl_i/h =[\frac14,\frac34]^T$, $\alpha_{in}=0$, $|\ain|=a/4$,  $\angle \h\eta_{101} = 10$\textdegree, $\angle \h\eta_{102} = -20$\textdegree, $\angle \h\eta_{201} = 30$\textdegree, $\angle \h\eta_{202} = -60$\textdegree;
		$t=0$.
	}%
	\label{fig:validation:w2wb2}%
\end{figure}

Validation of lateral and horizontal boundary conditions \eqref{eq:system_lin} is presented in \autoref{fig:validation:w2wb2}, showing the equation term values in physical space for two flaps moving at two frequencies. 
Results are scaled with and independent of water depth $h$ and characteristic steepness $ka$.
Example parameters and flap angles are chosen somewhat arbitrarily with the intention of displaying representative profile features. 
The cross-terms $P\oo2$, $R\oo2$ and $Q\oo2$ are computed in physical space, directly from \eqref{eq:defPRQ}, using the first-order amplitudes $\h\phi\ijn\oo1$, $\h\eta\ijn\oo1$ and $\hX_{in}\oo1$. 
Uniform weights  $w_i=1/N$  were adopted for \eqref{eq:hX_weight}.

\begin{figure}[H]%
	\centering
	\subfloat[second-order lateral boundary condition.]{\includegraphics[width=.5\columnwidth]{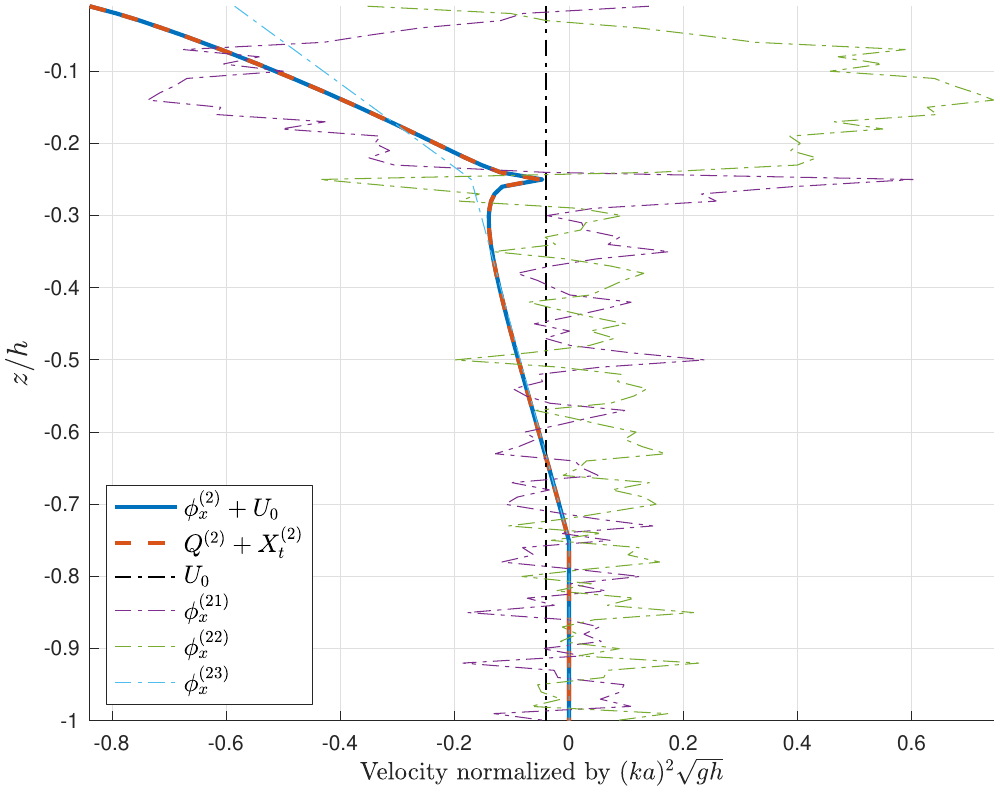}}%
	\subfloat[second-order kinematic condition.]{\includegraphics[width=.5\columnwidth]{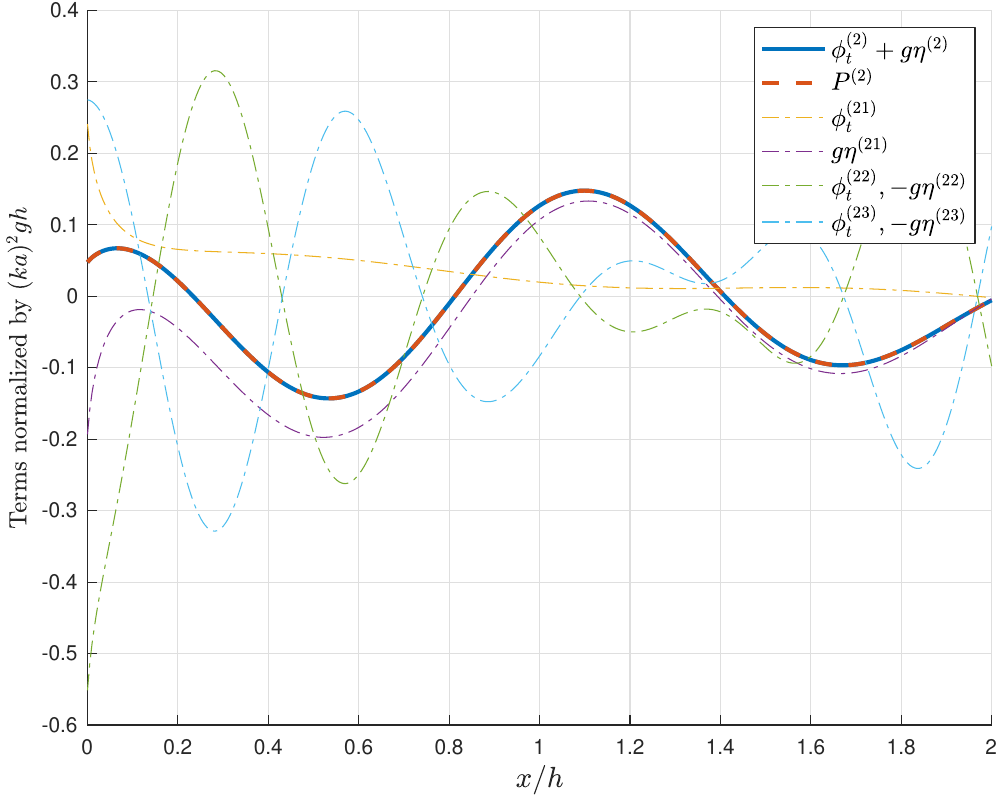}}%
	\caption{Repetition of benchmark \eqref{fig:validation:w2wb2} with directional waves $\alpha_{in}=30$\textdegree.}%
	\label{fig:validation:w2wb2_alpha}%
\end{figure}

The  current component $U_0$ amounts to a substantial part of the wall velocity in \autoref{fig:validation:w2wb2:QO2}.
Indeed, such back-flows can be observed visually in wave flumes with the aid of reflective particles and a laser. 
The free (22) and (23) wave components satisfy the homogeneous horizontal boundary conditions by design. These are plotted term-wise  in \autoref{fig:validation:w2wb2:PO2} and \ref{fig:validation:w2wb2:RO2}. 
A final validation with directional waves, $\alpha=30\mdeg$, is included in \autoref{fig:validation:w2wb2_alpha}.
Individual potential components here become more noisy, but this does not affect the summed potential.

The remaining boundary conditions, as well as the Laplace equation, are satisfied by construction. 
The multi-hinge wavemaker problem is thereby solved to second order.
A brief experimental validation of the theory at hand has been presented in \citet{fouques2022OMAE_multihinge}, and similar experiments are presented in the section that follows.

\section{\rev{Characteristics of multi-hinged wavemakers}}
\label{sec:optimize}

\rev{
	Multiple wavemaker hinges provide flexibility for wave generation and second-order correction. 
   This section explores the characteristics of multi-hinged wavemakers and how combination of paddle motions influence spurious wave generation. 
   Particular focus is given to single-harmonic paddle motions designed to completely remove second-order spurious waves.
   \citet{fouques2022OMAE_multihinge} reported that conversional second order correction, which introduce paddle motion at the double-frequency, can itself generate significant spurious waves at third order.
   Experiments are conducted to determine whether single-harmonic correction improves  overall wave quality compared to double-harmonic correction.
   Wave fields containing multiple frequencies are then considered, followed by examples of completely flexible wavemakers.
}

\subsection{\rev{Monochromatic paddle motions that suppress spurious waves}}
\rev{
Consider monochromatic motion ($n=\pm1$) of a multi-hinged wavemaker paddle that generates a progressive wave of amplitude $\A$ and wavenumber $k=k_{01}$.
Without resorting to double-frequency motions, is there a combination of  paddle strokes that will eliminate the spurious wave?
What do such motions look like?
}

The problem can be written as the following nonlinear system of two complex equations:
\begin{align}
\bigg\{ \h\eta_{0nm}\oo{22}=0,\quad \sum_{i=1}^N\h\eta_{i0n}\oo1 = \An\bigg\},
\label{eq:optSystMono}
\end{align}
with $n=m=1$ and $\An = \A$ for regular waves.
Each hinge provides one complex degree of freedom in $\h\eta_{i01}\oo1$ or, equivalently, $\hX_{i1}\oo1$.
Accordingly, $N=2$ (a double-hinge flap or a piston--flap wavemaker) should be sufficient to eliminate dispersive waves with monochromatic wavemaker motion. 
Additional hinges $N>2$ provide additional degrees of freedom that reduce the wavemaker draught. 
\\

\begin{figure}[H]%
\centering
\subfloat[Wave amplirtude.]{\includegraphics[width=.5\columnwidth]{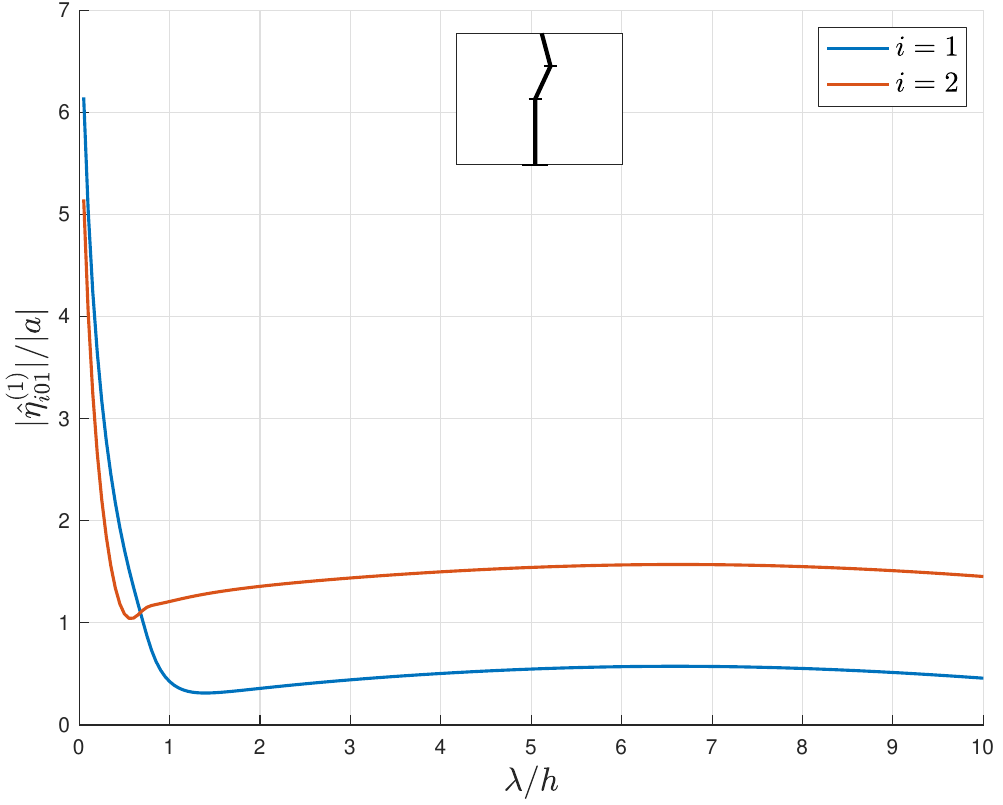}}%
\subfloat[Wave phase.]{\includegraphics[width=.5\columnwidth]{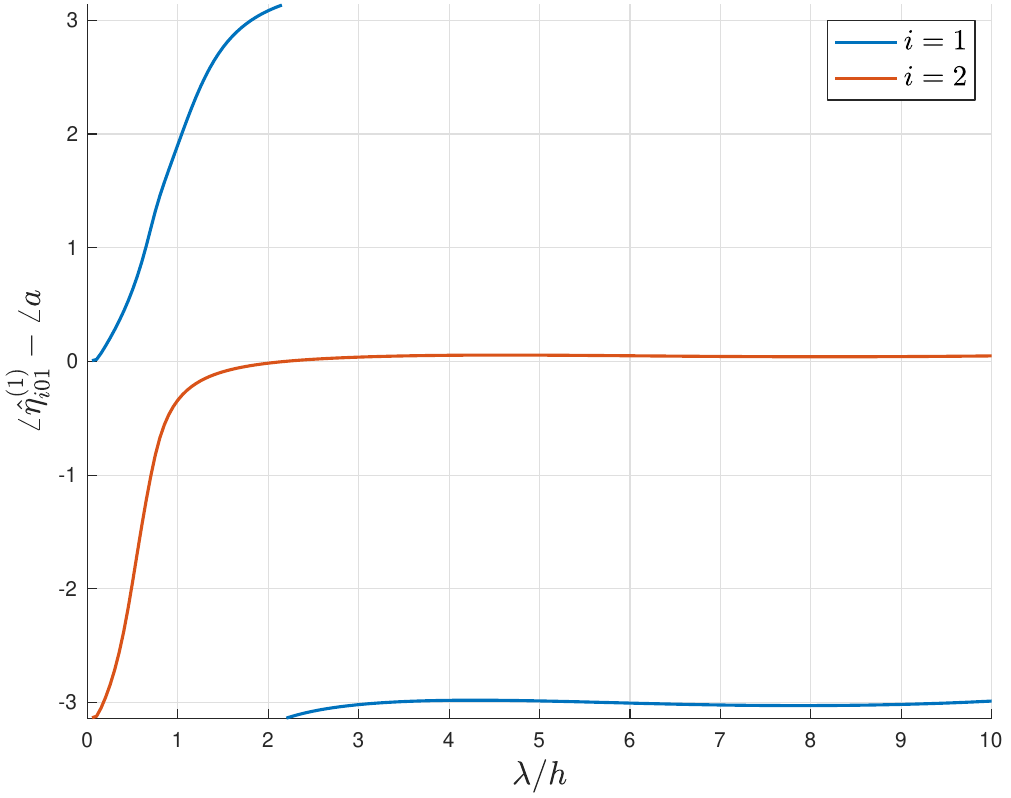}}
\caption{
	\rev{Monochromatic double-hinged wavemaker motions that cancel second-order spurious waves.
	Respective amplitude and angle of the linear wave generated by each paddle.}
	Hinge depth $\wbl_i/h=[\frac14,\frac12]$ as illustrated in the small inner panel in the left plot.
}%
\label{fig:wbOptwb14_12}%
\end{figure}

\begin{figure}[H]%
\centering
\subfloat[Wave amplirtude.]{\includegraphics[width=.5\columnwidth]{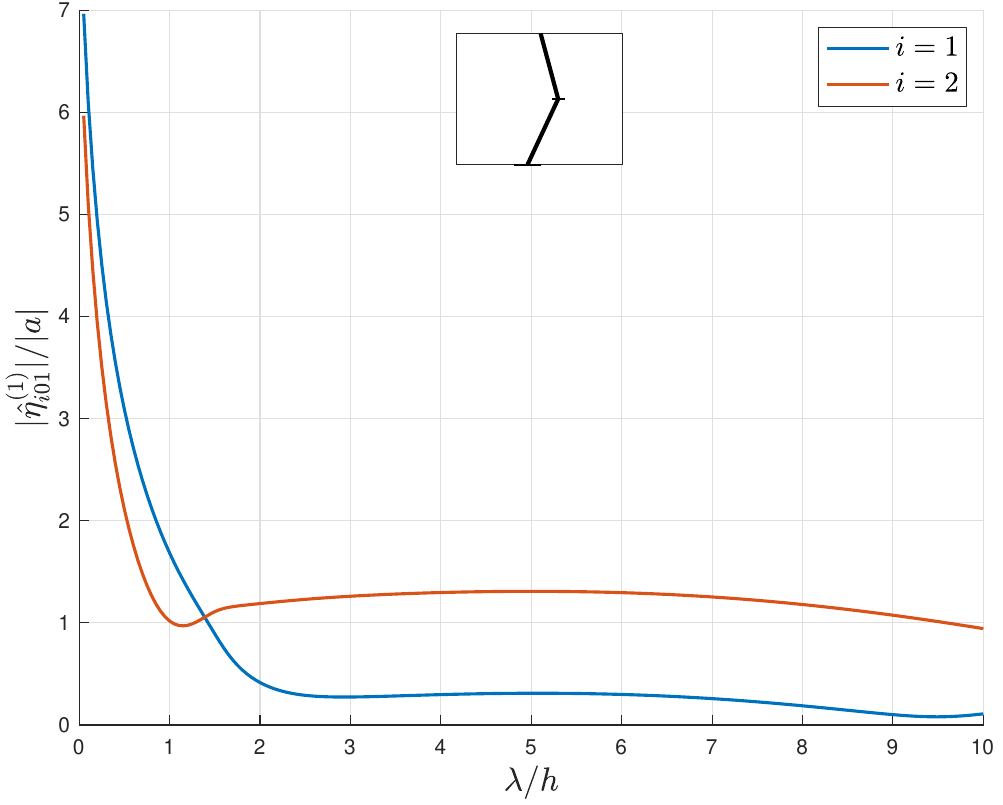}}%
\subfloat[Wave phase.]{\includegraphics[width=.5\columnwidth]{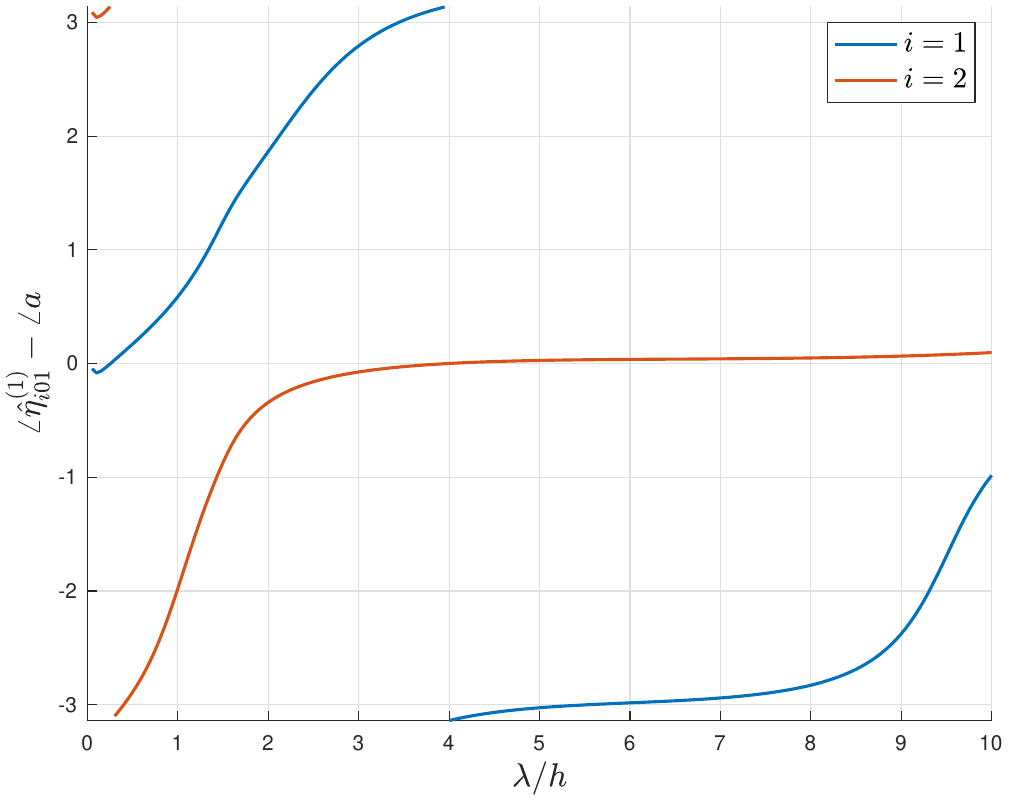}}
\caption{
	As \autoref{fig:wbOptwb14_12}, with  hinge depth $\wbl_i/h=[\frac12,1]$.
}%
\label{fig:wbOptwb12_11}%
\end{figure}

\begin{figure}[H]%
\centering
\subfloat[Wave amplirtude.]{\includegraphics[width=.5\columnwidth]{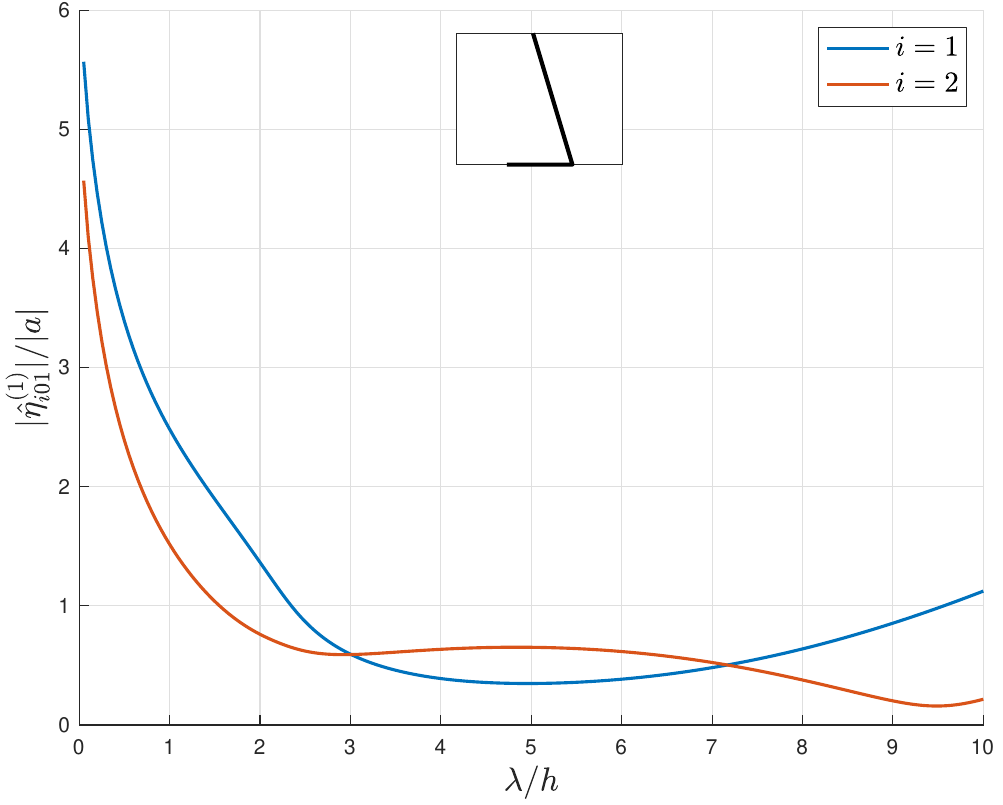}}%
\subfloat[Wave phase.]{\includegraphics[width=.5\columnwidth]{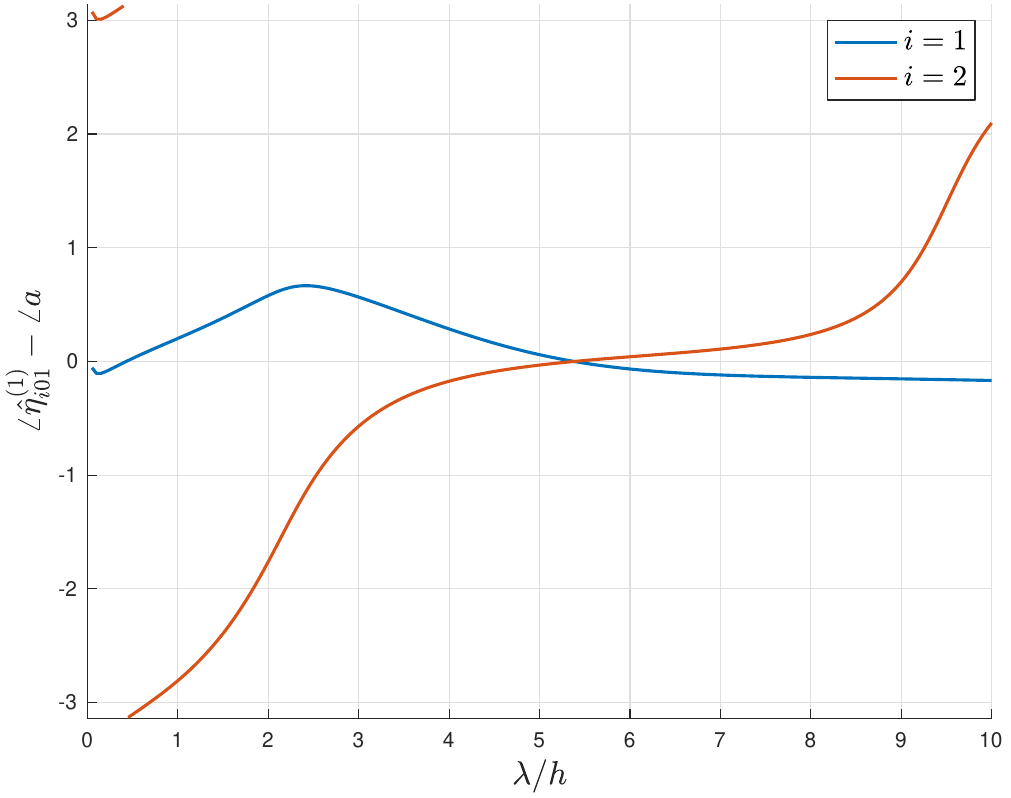}\label{fig:wbOptwb1_inf:ang}}%
\caption{As \autoref{fig:wbOptwb14_12}, with a deep hinged  $\wbl_1=h$ atop a piston  $\wbl_2\to\infty$.
}%
\label{fig:wbOptwb1_inf}%
\end{figure}

\begin{figure}[H]%
\centering
\subfloat[Wave amplirtude.]{\includegraphics[width=.5\columnwidth]{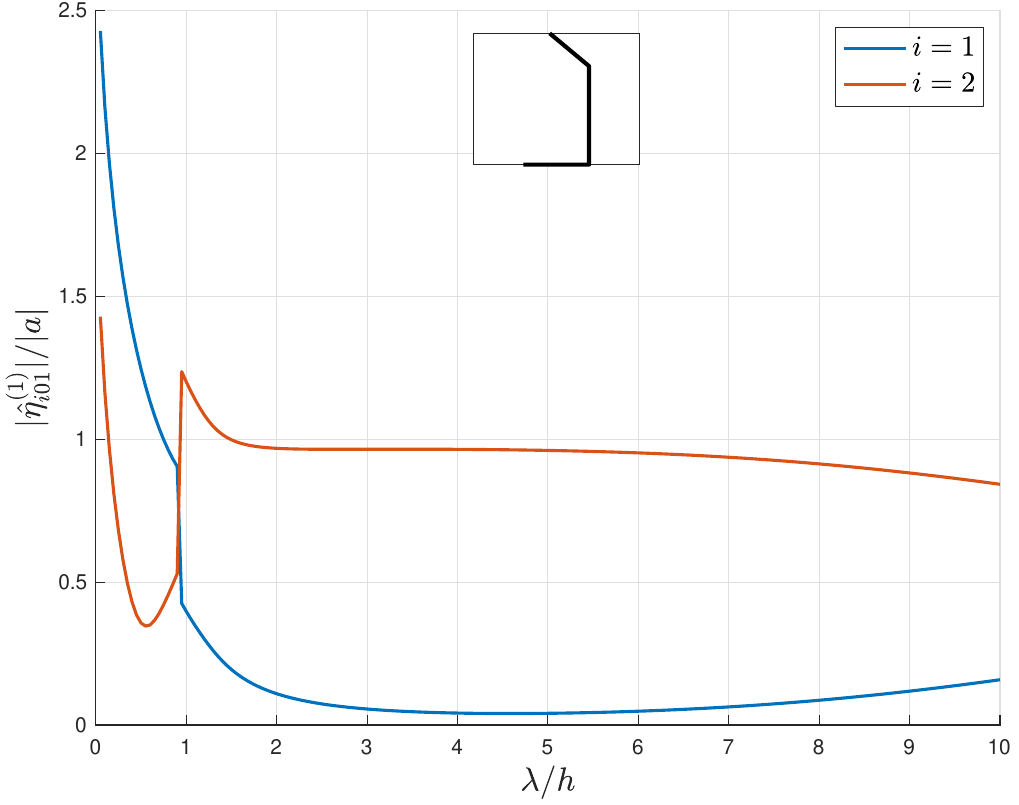}}%
\subfloat[Wave phase.]{\includegraphics[width=.5\columnwidth]{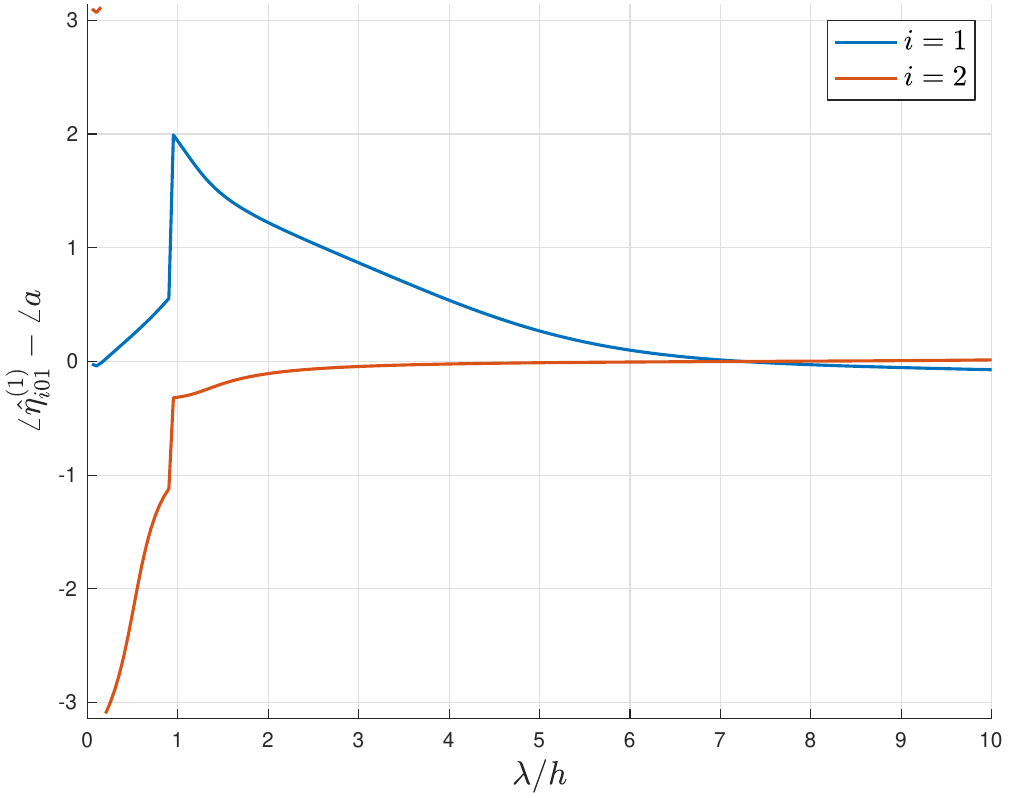}}%
\caption{As \autoref{fig:wbOptwb14_12}, with a shallow hinge $\wbl_1/h=\frac14$ atop a piston  $\wbl_2\to\infty$.
}%
\label{fig:wbOptwb14_inf}%
\end{figure}

Double-hinged paddle motions that result from solving  \eqref{eq:optSystMono}  are plotted in \autoref{fig:wbOptwb14_12} to \ref{fig:wbOptwb14_inf}.
Double-flap wavemakers mostly in near opposite phase, with
the largest horizontal displacement  occurring underneath the waterline.
\\

\begin{figure}[H]%
\centering
\subfloat[Case in \autoref{fig:wbOptwb14_12}.]{\includegraphics[width=.5\columnwidth]{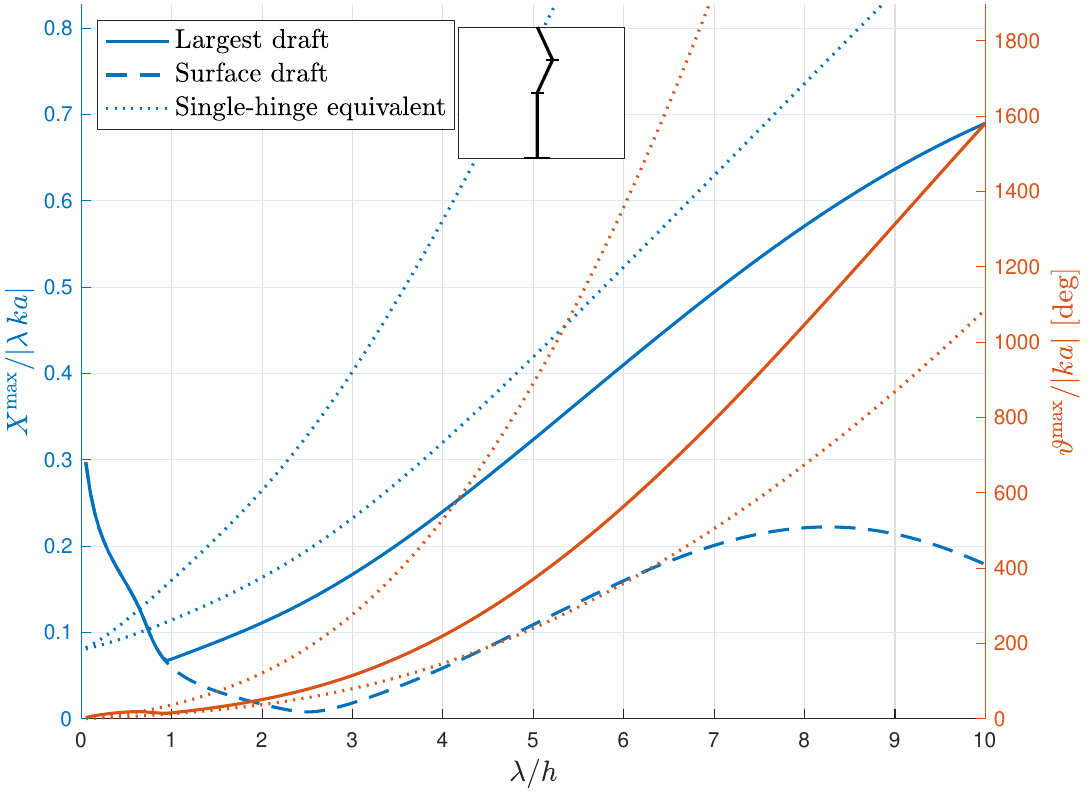}}%
\subfloat[Case in \autoref{fig:wbOptwb12_11}.]{\includegraphics[width=.5\columnwidth]{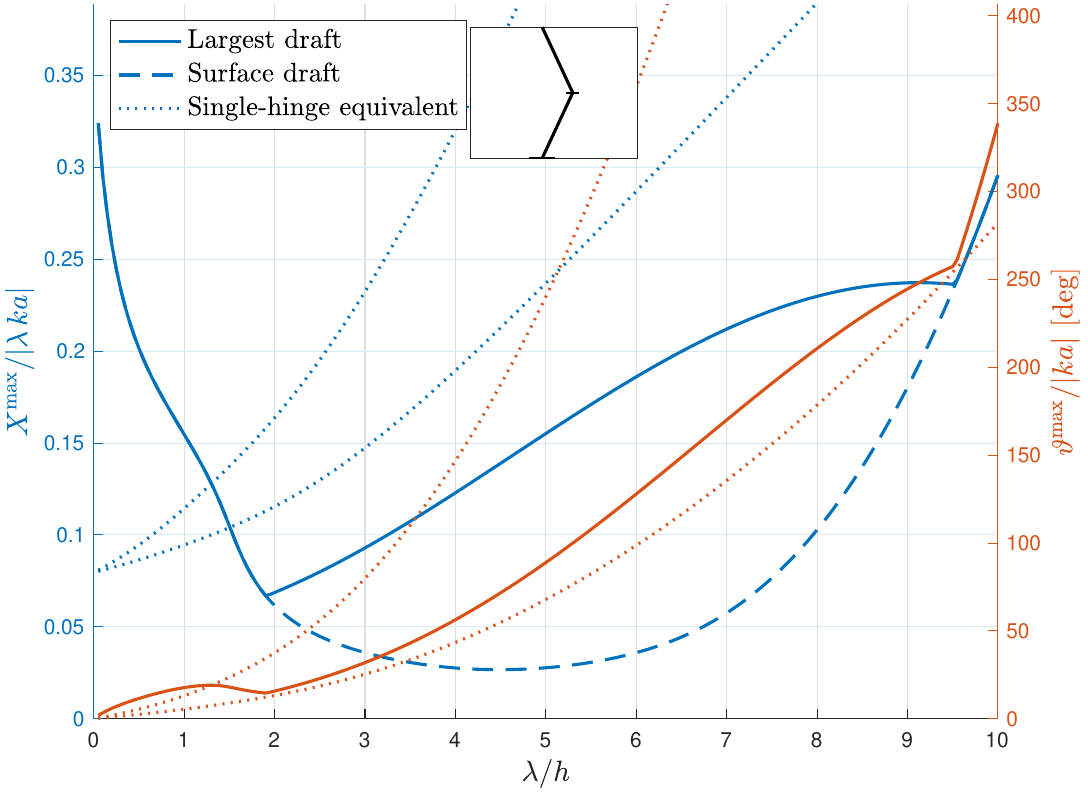}}\\
\subfloat[Case in \autoref{fig:wbOptwb1_inf}.]{\includegraphics[width=.5\columnwidth]{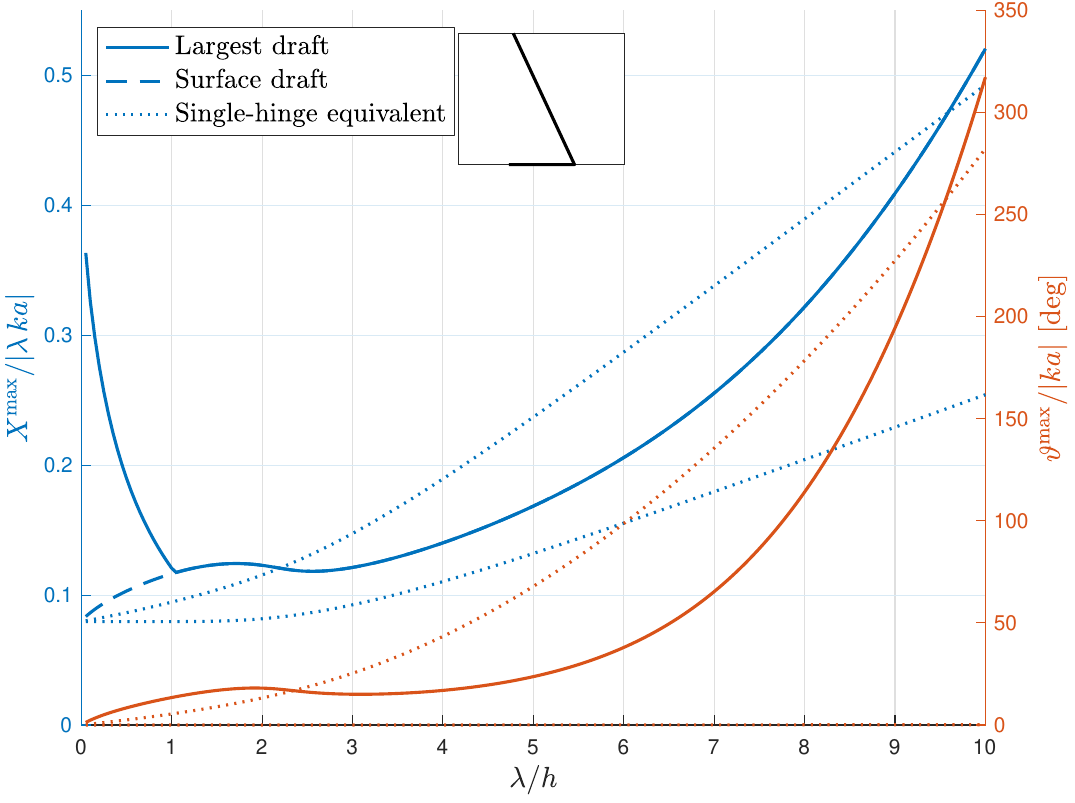}}%
\subfloat[Case in \autoref{fig:wbOptwb14_inf}.]{\includegraphics[width=.5\columnwidth]{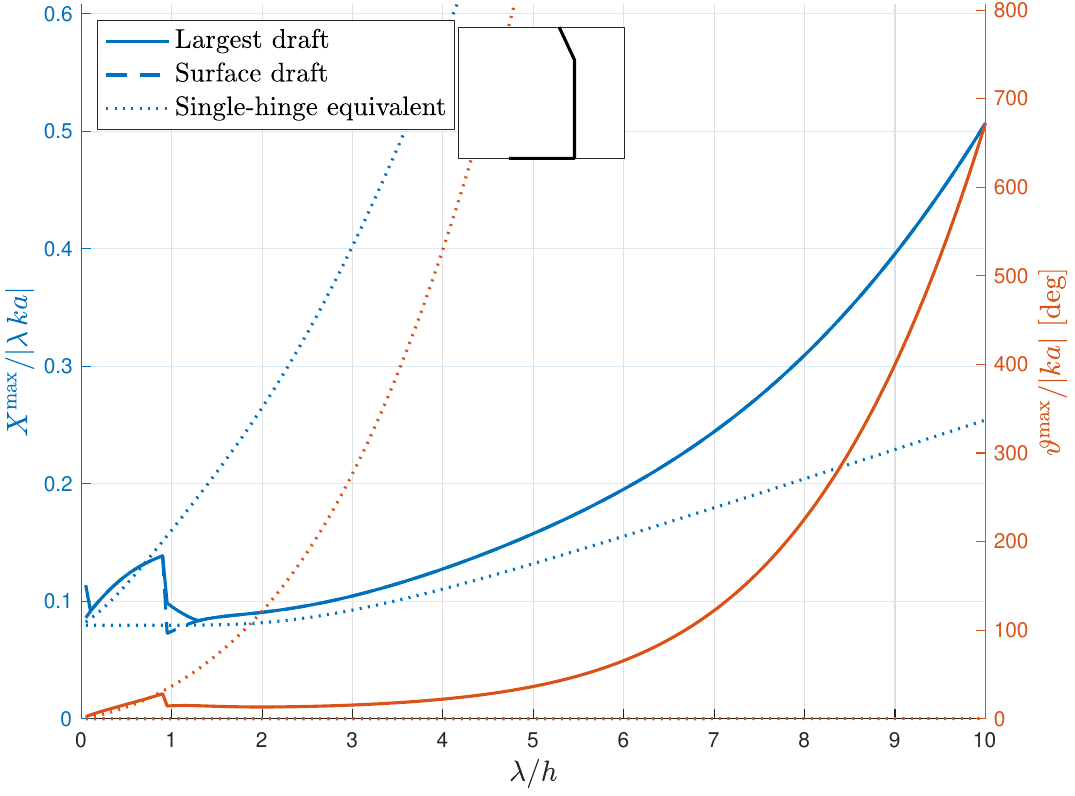}}%
\caption{Largest paddle displacement $X\^{max}$ and cumulative paddle angle $\vartheta\^{max}$ expressed by \eqref{eq:Xmax} for wavemakers presented above.
A dashed line shows the paddle displacement at the water surface. 
Dotted lines show the equivalent displacement if the same wave is generated with only one of the paddle, 
the upper and lower  line representing  the upper  and lower paddle, respectively.
}
\label{fig:wbhX}%
\end{figure}

The present theory assumes small paddle displacements $X(y,z,t)$ and hinge angle $\theta_i(t)$. 
Paddle motion should be restricted in these quantities for the sake of accuracy and mechanical capability.
As just shown, with multiple hinge points, the largest displacements may occur underneath the waterline.
An expression for the largest paddle displacement and angle is obtained using the cumulative hinge angle \eqref{eq:bX_lin} in \appendixref{sec:BC_x}, which to linear order yields
\begin{align}
\vartheta\^{max} &= \max_{\tilde N} \Big|\sum_{i=1}^{\tilde N} \h\theta_{i1}\oo1 \Big|,
&
X\^{max} &= \max_{\tilde N} \Big|\sum_{i=1}^{\tilde N} (\wbl_i - \wbl_{i+1}) \h\theta_{i1}\oo1 \Big|;\; \wbl_{N+1}=0.
\label{eq:Xmax}
\end{align}
Hinge angles are defined similar to the displacement, 
with
	$\theta_i\oo1 = \frac12\sum_n \h\theta_i\oo1\ee^{\ii (\wn t - \kyn y)}$, $\h\theta_{in}\oo1 = \hX_{in}\oo1/\wbl_i$.
These measures are plotted in   \autoref{fig:wbhX} for the same paddle motion shown in \autoref{fig:wbOptwb14_12}--\ref{fig:wbOptwb14_inf}, and are scaled by wave steepness and wavelength.
\rev{Also shown in the figure is the corresponding displacement and hinge angle needed to generate the same linear wave moving only one of the hinges. 
Surprisingly, less stroke is needed to generate medium and long waves  with the double-hinged flaps compared to using either flap alone.
This is due to the opposing phases,
and the displacement at the water surface is seen to be even less. 
In terms of angles, the cumulative angle will lie in between the equivalent angles of the upper and lower flaps.
With combined piston--flap wavemakers, the piston alone usually requires the smallest draft.
 }

\begin{figure}[H]%
 	\centering
 	\includegraphics[width=.5\columnwidth]{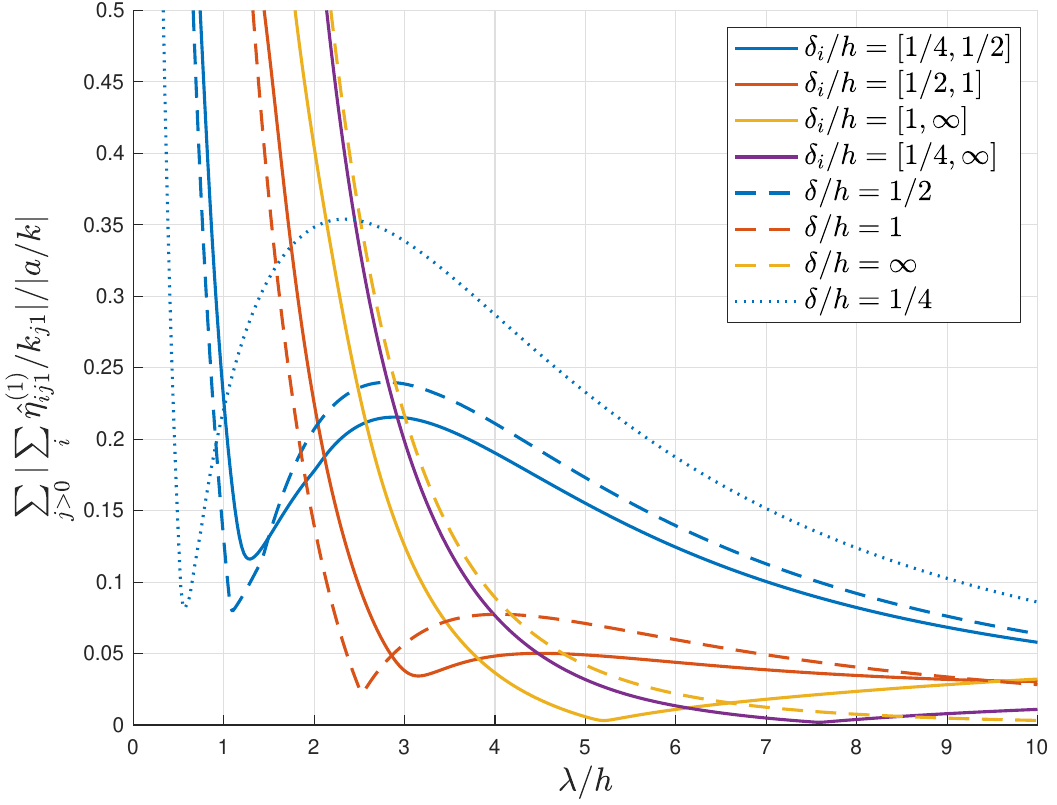}%
 	\caption{
 		Weighted measure of the intensity of the evanescent near field, $\sum_{j>0}|\sum_i\h\eta_{ij1}\oo1/k_{j1}|/|a/k|$, 
 		comparing the intensity of double-hinge single harmonic wave generation (solid line) to wave single-hinged wave generation (dashed).
 	}%
 	\label{fig:evO1}%
 \end{figure}

\rev{Before the introduction of nonlinear wavemaker theory, it was sometimes assumed that wave quality could be related to the intensity of the evanescent near field. 
	A measure of this intensity,  $\sum_{j>0}|\sum_i\h\eta_{ij1}\oo1/k_{j1}|$, is scaled by steepness an plotted in \autoref{fig:evO1} for the double-hinged, single harmonic wavemaker motions that  cancel  spurious waves. 
	These are compared to near fields generated when moving only one paddle, as would be the case if wave correction was enforced through double-harmonic motion.
	The plot indicates no significant relationship between near field intensity and the special double hinged motions. 
	Rather, it is observed that the near field is small when the hinge depth is about 1/3 of the wavelength, which, if desired,  can be better accommodated with multi-hinged paddles.
}
\\

\begin{figure}[H]%
\centering
\subfloat[Deep water range]{\includegraphics[width=.5\columnwidth]{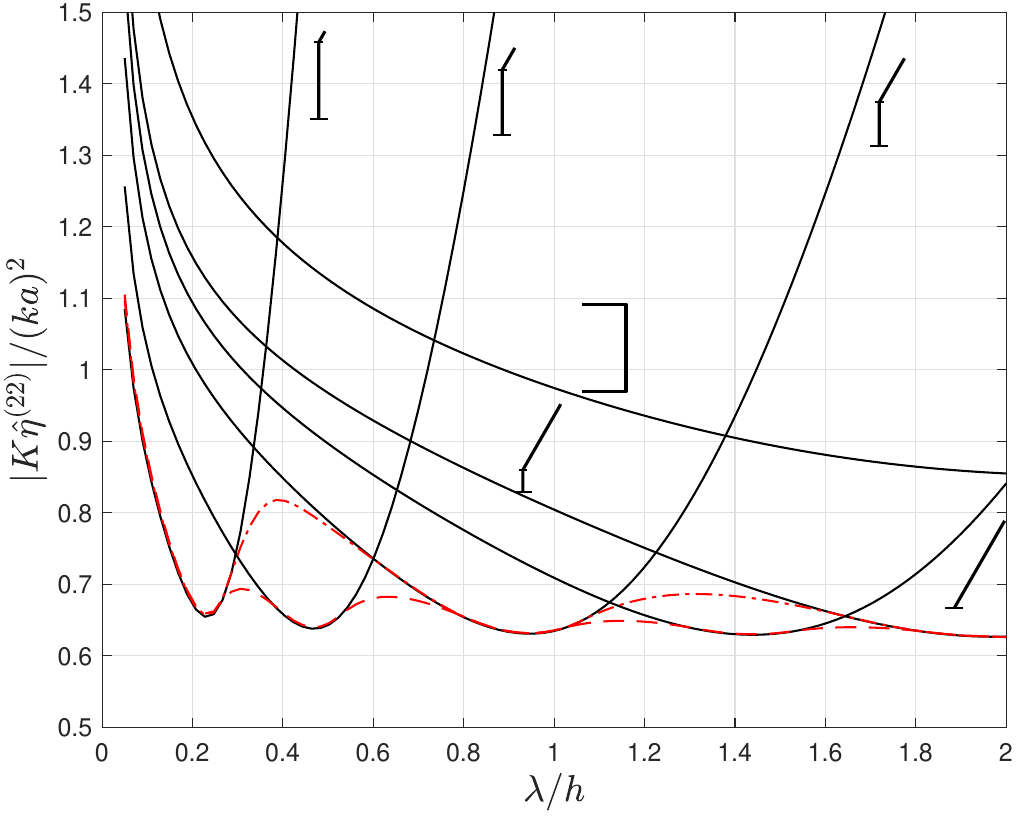}}%
\subfloat[Full  range]{\includegraphics[width=.5\columnwidth]{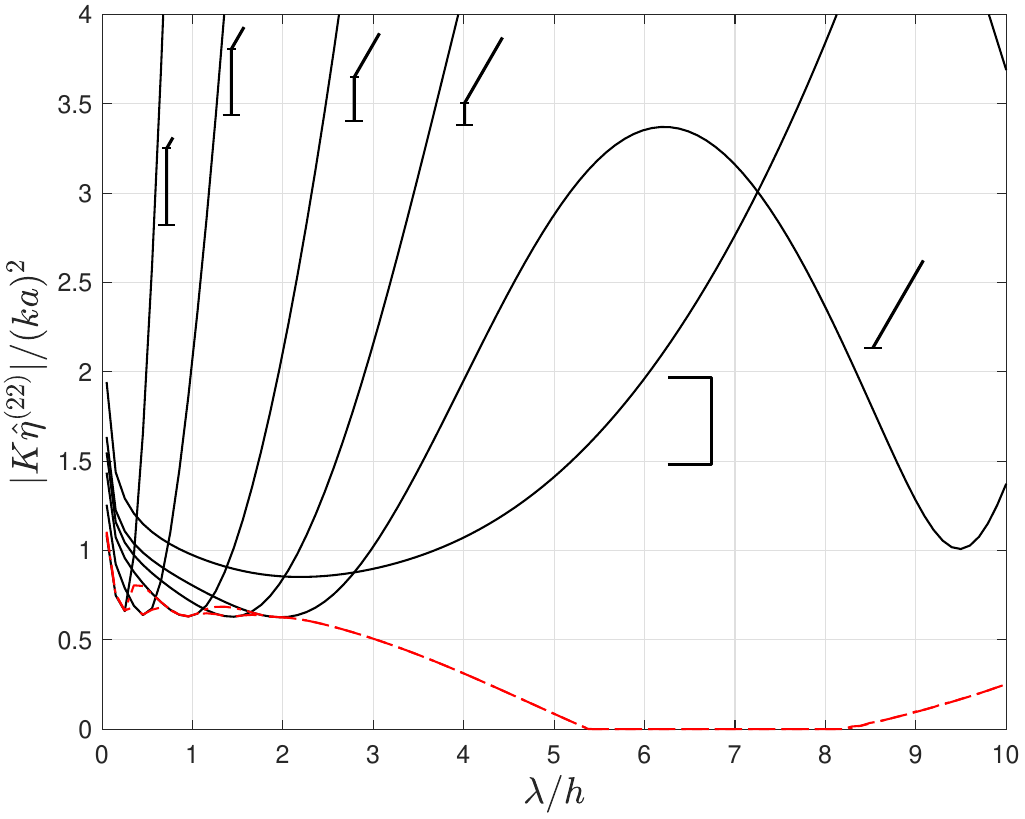}}%
\caption{
Spurious wave steepness, scaled by primary wave steepness, for singe-hinged wavemakers and double-hinged wavemakers forced to move both paddles in phase. 
Paddle depths are $\wbl/h \in\{\frac18,\frac14,\frac12,\frac34,1, \infty\}$ as visualized with small sketches.
Red lines represent double-hinged wavemakers combining two hinge depths. 
}%
\label{fig:wbAmplitude:heta21__22}%
\end{figure}

\rev{An assumption made in the past about double-hinged wavemakers is that the paddles should be moved in phase \citep{fouques2022OMAE_multihinge}.
Although our insight is limited to second-order theory, the study just presented shows that the lowest spurious wave generation occurs when the paddles are in near-opposite phase, contradicting the assumption.
Minimising spurious wave under the restriction that paddles move in phase leads to the conclusion drawn in  \citep{fouques2022OMAE_multihinge}, that the near `optimal' motion is moving only one paddle at a time, depending on wavelength. 
\autoref{fig:wbAmplitude:heta21__22}  illustrates this, where curves of minimum free-harmonic intensity overlap those of single-hinged  wavemakers across most of the plotted range.
The figure also shows that the lowest spurious wave generation occurs when the hinge depth is about half a wavelength.
}

\subsection{Experimental validation of single-harmonic wave correction}
\label{sec:experiment}
\begin{figure}[H]%
	\centering
	\subfloat[Wavemaker]{\includegraphics[width=.5\columnwidth]{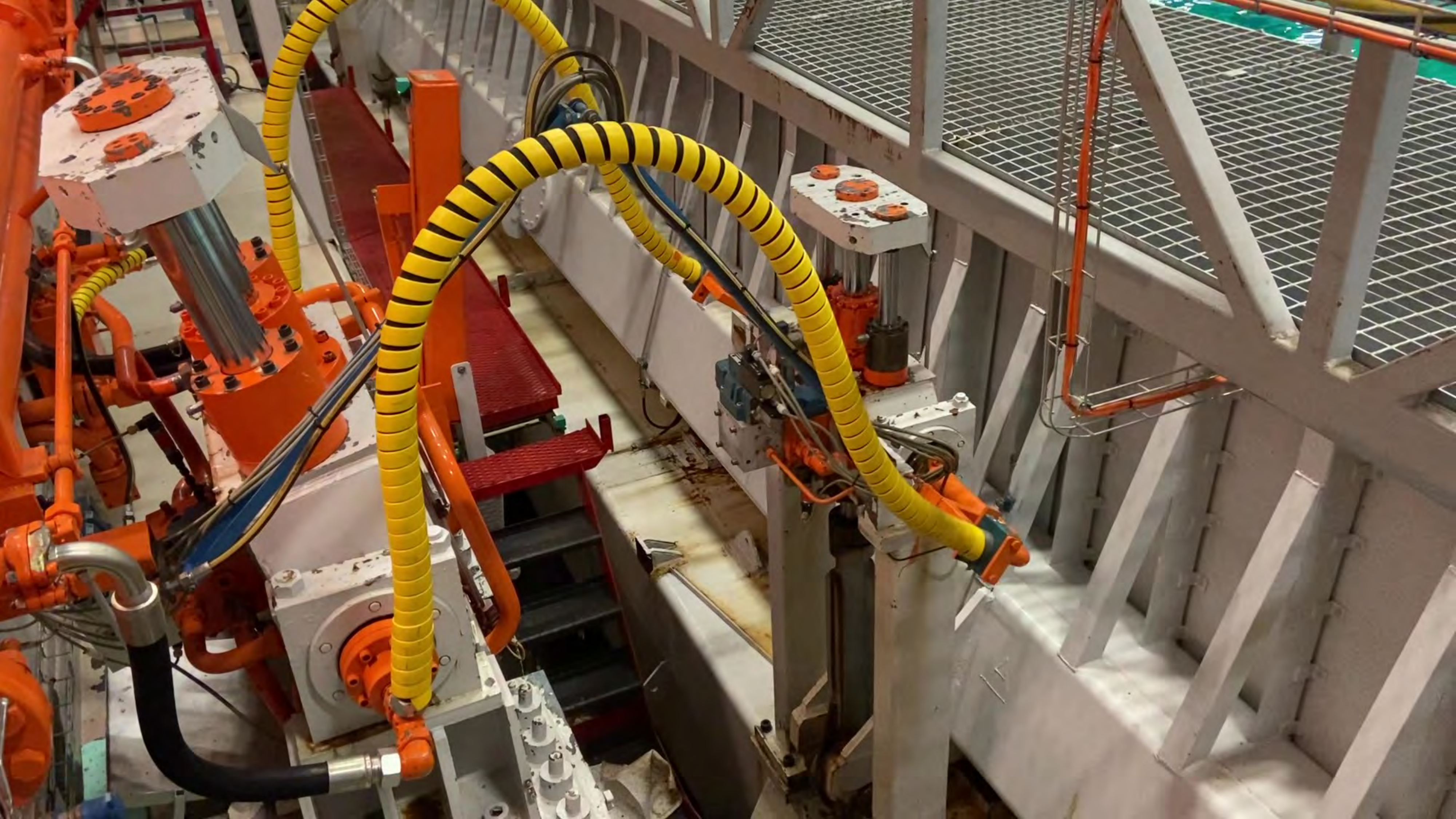}\label{fig:BM2photo}}%
	\quad
	\subfloat[Resistance wave gauges]{\quad\includegraphics[width=.35\columnwidth]{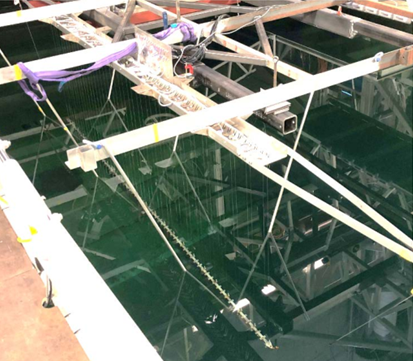}\quad\label{fig:harp}}%
	\caption{Experimental equipment.}%
\end{figure}

The SINTEF Ocean Basin laboratory is furnished with the double hinged wavemaker sketched in \autoref{fig:BM2} and photographed in \autoref{fig:BM2photo}.
This wavemaker is not ideal since it has  protruding parts along its flap surface, but it serves for validation.
Hinges are located at depths $\wbl_1=1.05$ and $\wbl_1=2.62$ meters, and the water depth is 5.0 meters. Only deep water periods, $T=1.50$, 1.75, 2.00, 2.25 and 2.50 seconds, are tested to avoid  disturbances from a step in the bathymetry.
Wave steepness is kept around $ka\approx 0.2$.

An array of twenty-three equally spaced resistance type wave gauges, \autoref{fig:harp}, is used to decompose  wave  components.
Amplitude decomposition is performed by fitting the signals $\eta_l(t)$ from wave gauge $l$ to a linear model
\begin{align}
	2\eta_l(t) &= 
	\h\eta\_I\oo{1}\ee^{\ii\w t-\ii k x_i}
	+\h\eta\_R\oo{1}\ee^{\ii\w t+\ii k x_i}
	+\h\eta\_I\oo{21}\ee^{ 2\ii\w t-2 \ii k x_i}
	+\h\eta\_R\oo{21}\ee^{2\ii\w t+2\ii k x_i}
	\nonumber\\&
	+\h\eta\_I\oo{22}\ee^{2\ii\w t-\ii K x_i}
	+\h\eta\_R\oo{22}\ee^{2\ii\w t+\ii K x_i}
	+\h\eta\_I\oo{311}\ee^{3\ii\w t-3\ii k x_i}
	+\h\eta\_I\oo{312}\ee^{\ii \w t-\ii(K-k) x_i}
	\nonumber\\&
	+\h\eta\_I\oo{313}\ee^{3\ii\w t-\ii(K+k) x_i}
	+\h\eta\_I\oo{32}\ee^{3\ii\w t-\ii K\oo{3} x_i}
	+\varepsilon_i(t) +\text{c.c.}
	\label{eq:regressionModel}
\end{align}
using linear regression. Since the water is deep, we have $K\approx 4k$, $K\oo{3}\approx 9k$.

Monochromatic waves with and without double-harmonic correction are generated moving only the lower flap, 
along with  single-harmonic waves generated by moving both flaps in the motion optimised for eliminate spurious waves.
These are compared  in \autoref{fig:exp}.
The performance of the two correction methods are found to be comparable. 
A  discrepancy of principal amplitudes is seen in \autoref{fig:exp:O1}, originating from a difference in the applied wavemaker transfer functions.
Some second-order spurious waves remain also after correction (\autoref{fig:exp:O2}), likely caused by basin geometry, mechanical noise and measurement inaccuracy. 
A small amount of spurious waves is generated at third order, the magnitude of which appears similar in the two correction approaches (\autoref{fig:exp:O3}).
The noise level of these third-order components suggest that they have been contaminated, either by other disturbances in the basin, or by accuracy limitations of the amplitude decomposition,
and one can therefore only regard these results as indicative.
More discernable third-order waves can be generated  moving the upper wavemaker flap alone, but is does not compare fairly to double-hinged correction.  
More testing, experimental or numerical, is needed to  clearly determine whether single-harmonic wave correction can be beneficial to overall wave quality.

\def\wiii{.2\columnwidth}
\begin{figure}[H]%
	\centering
	\subfloat[{Primary harmonic $\eta\_I^{(1)}$\,[m].}]{ %
		\includegraphics[width=\wiii]{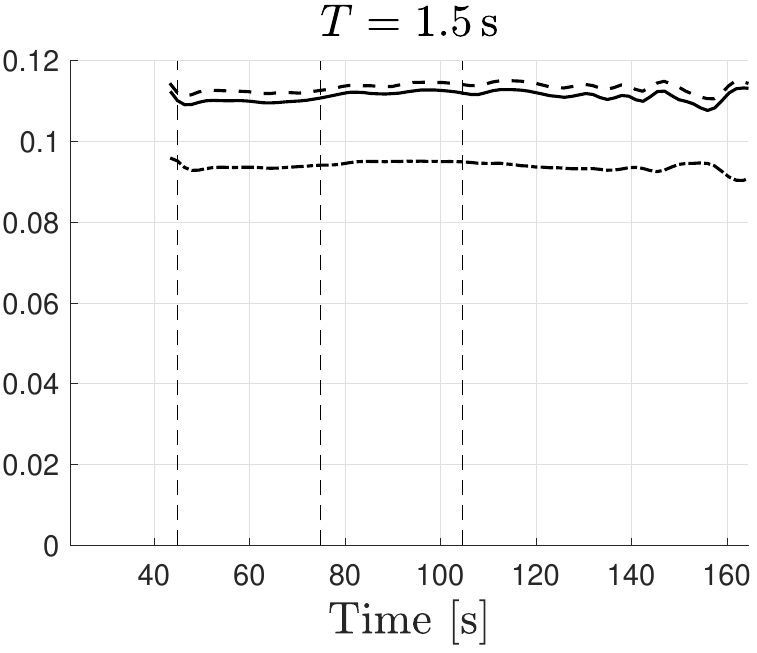}%
		\includegraphics[width=\wiii]{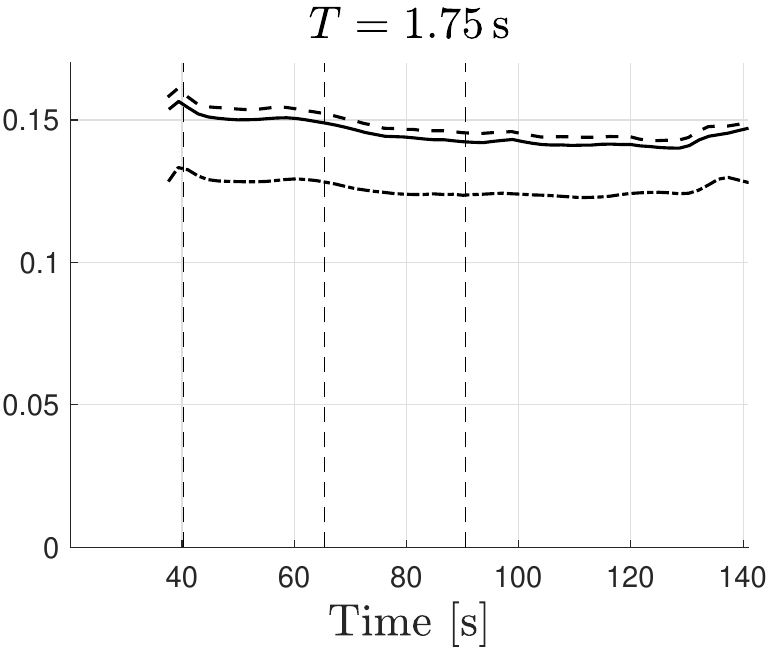}%
		\includegraphics[width=\wiii]{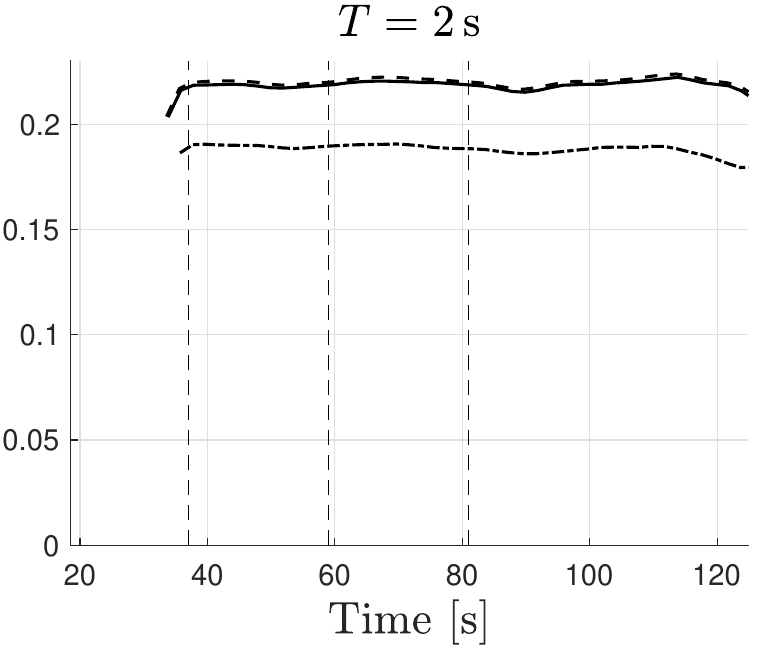}%
		\includegraphics[width=\wiii]{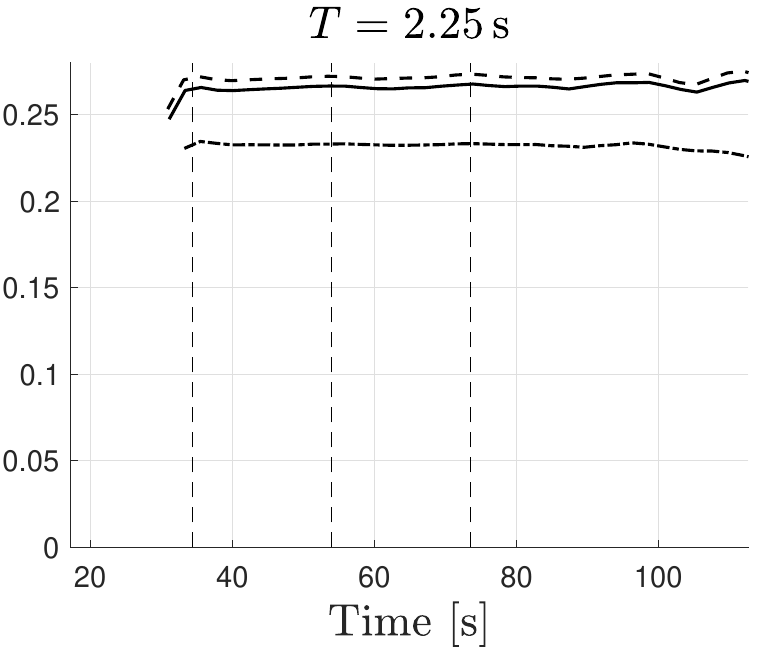}%
		\includegraphics[width=\wiii]{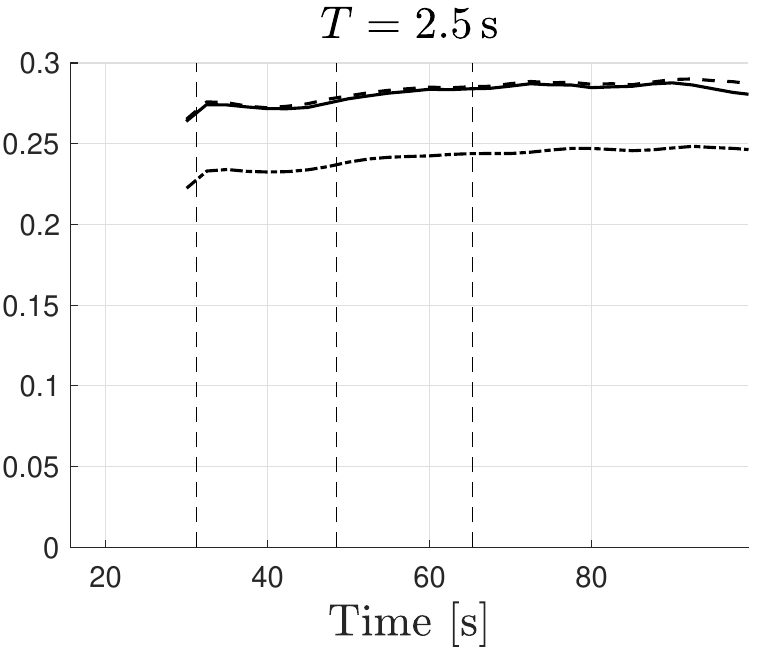}%
	\label{fig:exp:O1}}\\
	\subfloat[second-order spurious harmonic $\eta\_I^{(22)}/\langle\eta\_I^{(1)}\rangle^2$.]{ %
		\includegraphics[width=\wiii]{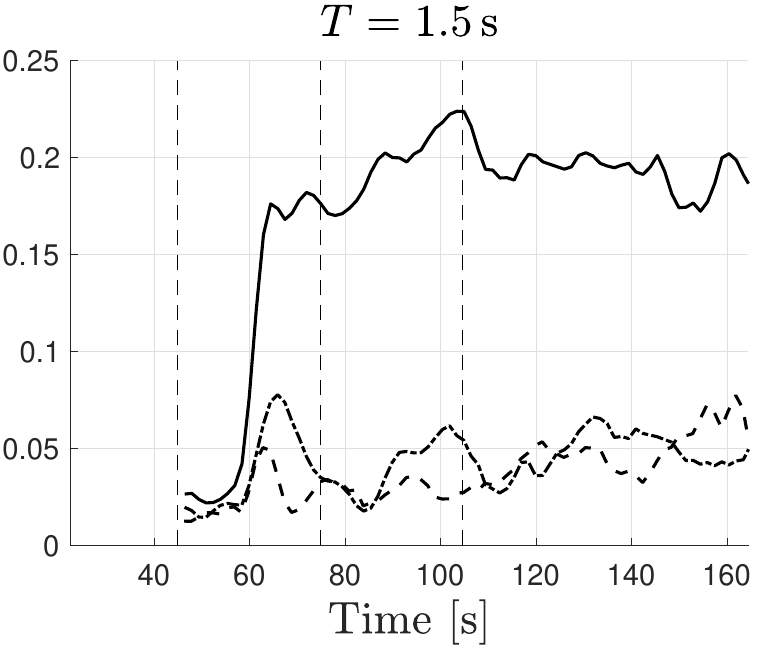}%
		\includegraphics[width=\wiii]{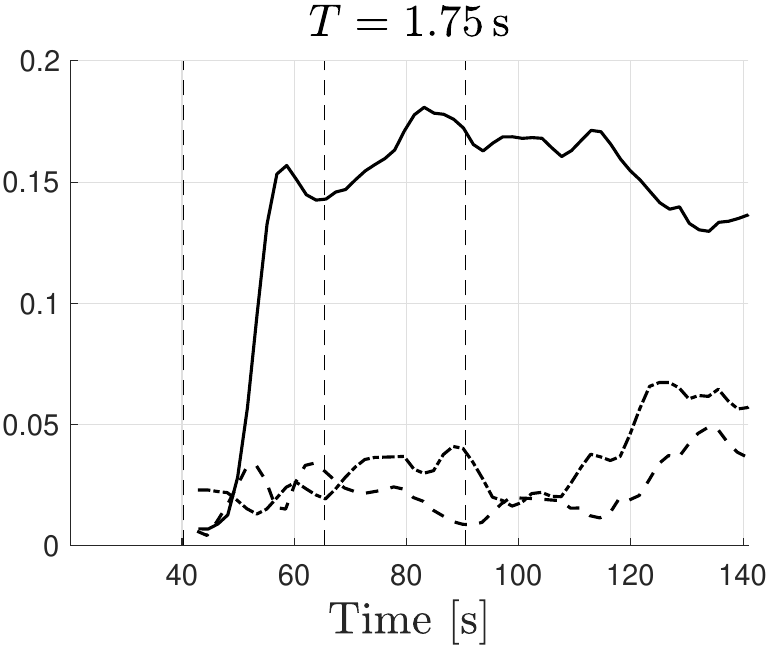}%
		\includegraphics[width=\wiii]{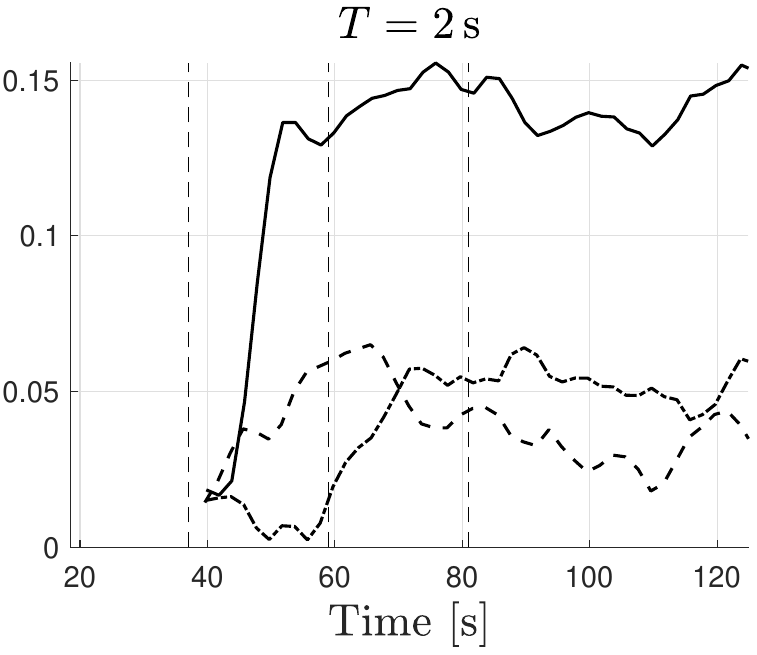}%
		\includegraphics[width=\wiii]{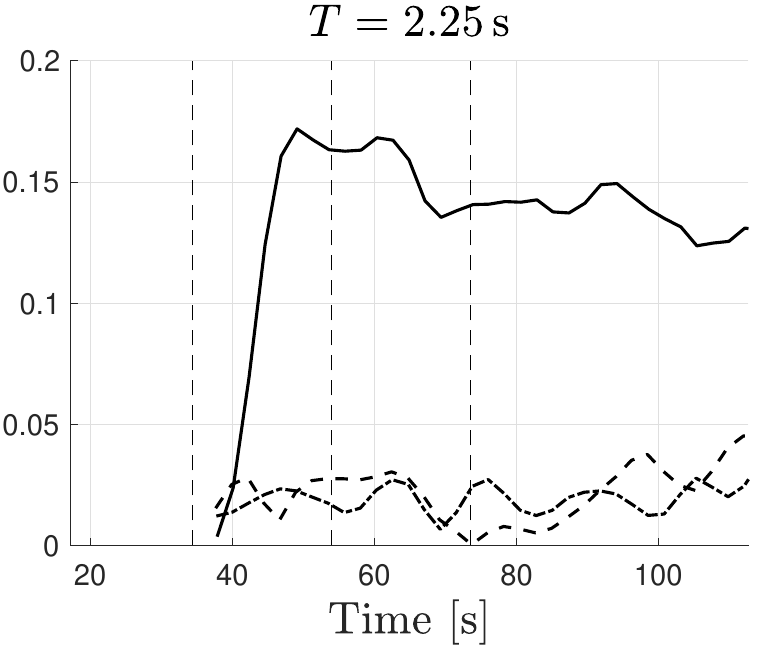}%
		\includegraphics[width=\wiii]{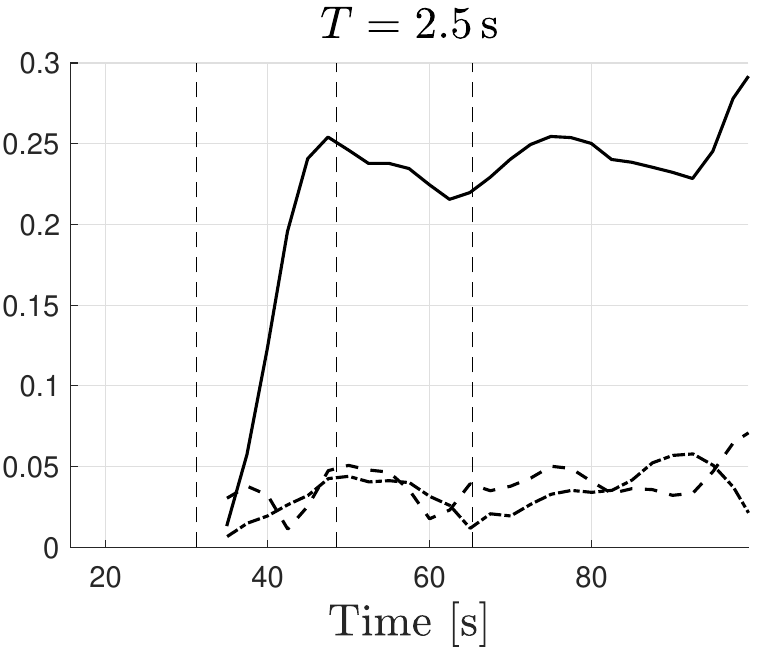}%
	\label{fig:exp:O2}}\\
	\subfloat[Third order spurious harmonic $\eta\_I^{(32)}/\langle\eta\_I^{(1)}\rangle^3$.]{ %
		\includegraphics[width=\wiii]{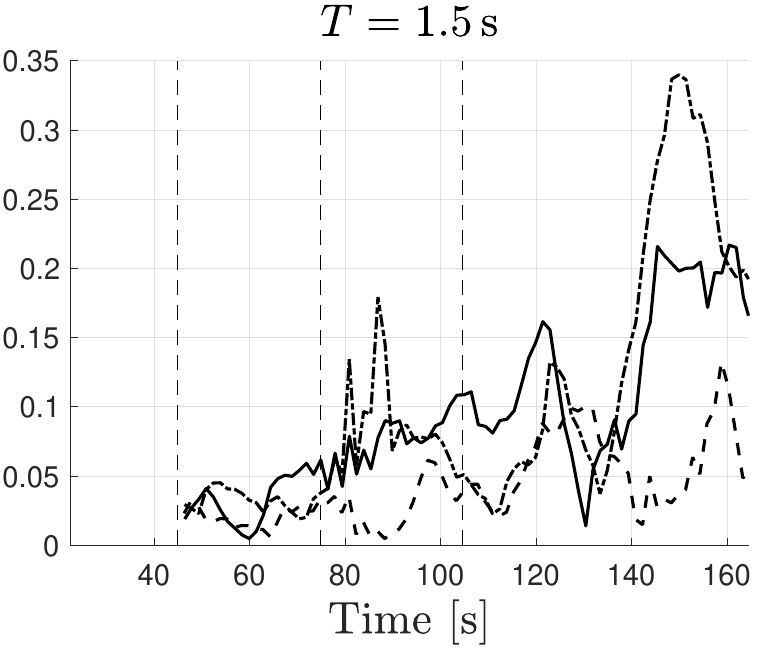}%
		\includegraphics[width=\wiii]{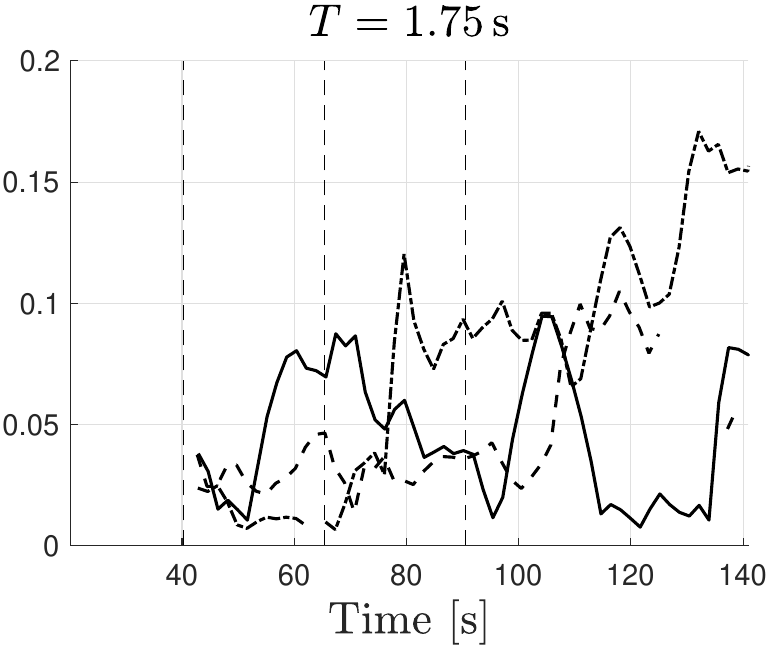}%
		\includegraphics[width=\wiii]{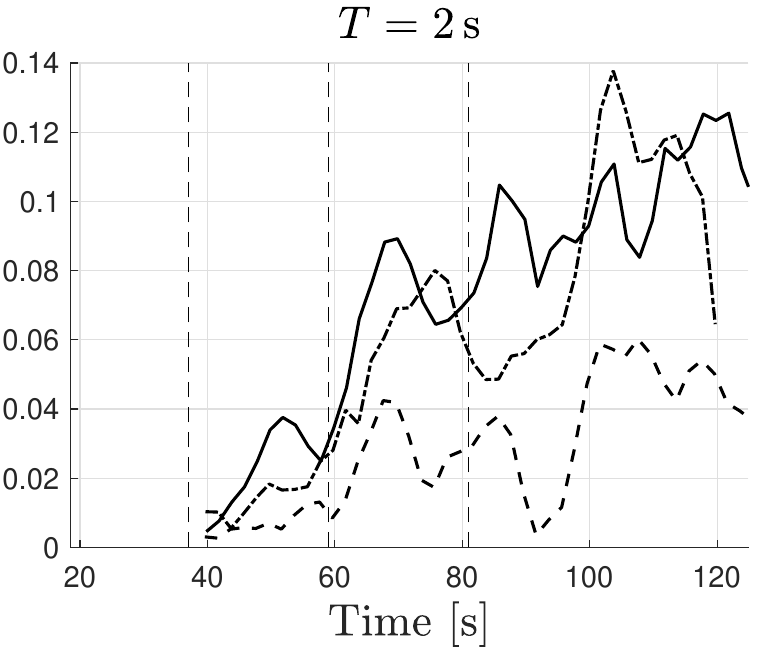}%
		\includegraphics[width=\wiii]{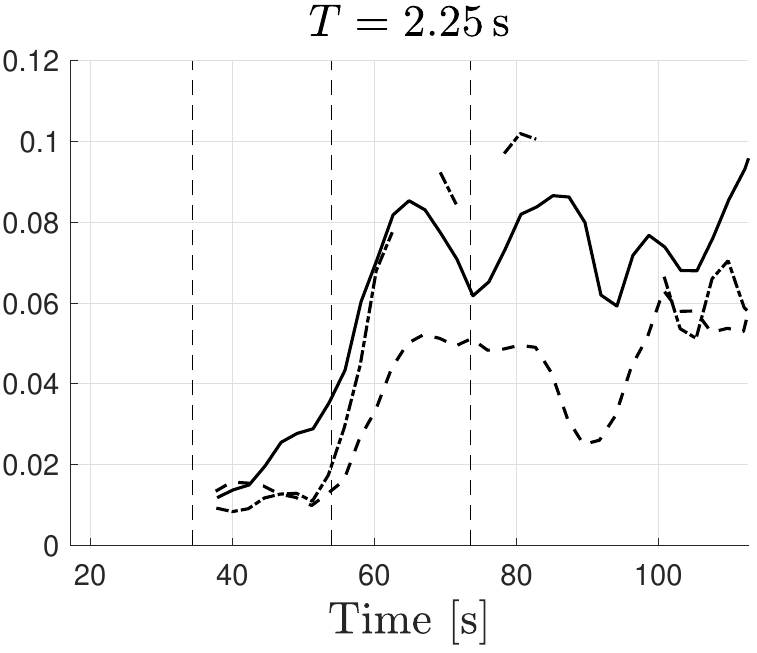}%
		\includegraphics[width=\wiii]{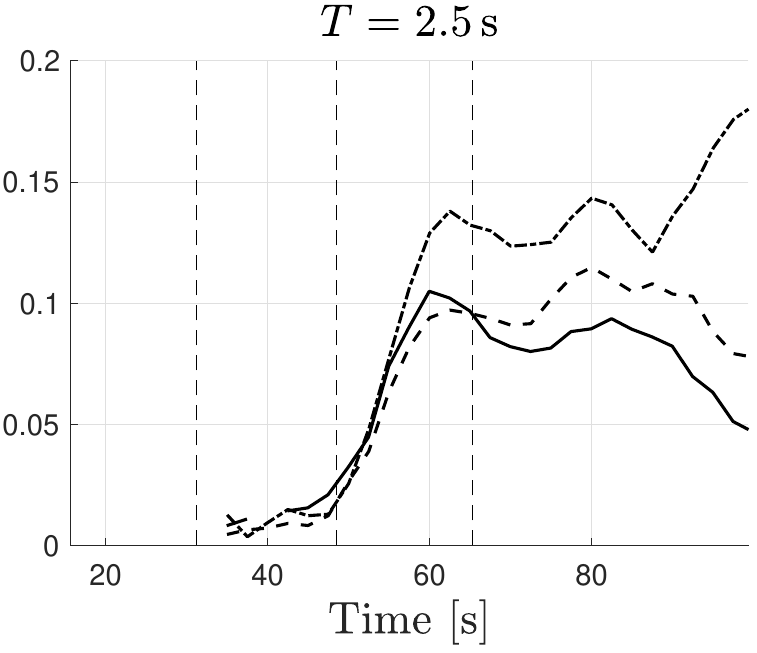}%
	\label{fig:exp:O3}}%
	\caption{Experimental amplitude components. Solid: monochromatic lower hinge motion (no wave correction); dashed: bi-chromatic lower hinge motion (conventional wave correction), dot--dashed: monochromatic double-hinge motion solving \eqref{eq:optSystMono}.
		Amplitudes fitted to model \eqref{eq:regressionModel} using linear regression over a three wave period sliding time window.
		Vertical lines indicate estimated arrival time of respective wave components based on group velocity. 
		Deep water and intermediate wave steepness $ka\approx 0.2$. }%
	\label{fig:exp}%
\end{figure}

\subsection{Briefly on  wavemaker motions of multiple frequencies }
Double-hinge systems have been shown capable of eliminating second-order regular spurious wave  under pure monochromatic motion. 
Is this transferable to wave fields of several frequencies?
To investigate, we assume that  $\Nw$ unique frequencies are imposed. 
Free to choose $\h\eta_{i0n}\oo1$, the number of complex degrees of freedom becomes $N\Nw$.
If none of the imposed frequencies are harmonics of each other, then the number of complex conditions to be satisfied is $\Nw^2$ for the first expression in \autoref{eq:optSystMono}, and  $\Nw$ for the second.
Accordingly, $\Nw(\Nw+1)$ equations must be satisfied to eliminate spurious waves, meaning  $N=\Nw+1$ hinges are needed to eliminate spurious waves without imposing higher-harmonic correction.

On the other hand, if frequencies $\wn$ are all harmonics, that is, $\wn= n\Delta\omega$ with $\Delta\omega$ constant, then the same harmonics will be repeated in the space of second-order interactions. 
Mode interactions then bunch into single-frequencies modes 
\begin{equation*}
\h\eta_{pq}\oo{22} = \sum_{n=q-\Nw}^{\Nw}\h\eta_{pn(q-n)}\oo{22}; \quad q=1,\ldots,2\Nw 
\end{equation*} 
for which  $\ww=q\Delta\omega$.
We are then left with $2\Nw$ equations first expression in \autoref{eq:optSystMono}, $\h\eta_{0q}\oo{22}=0$, totalling to $3\Nw$ conditions. 
$N=3$ is  a sufficient number of hinges for eliminating spurious waves in such a case, regardless of the number of frequencies.
\rev{Systems of this kind are indeed solvable, as demonstrated by  example in \autoref{tab:irr} for $\Nw=5$.
	Solving them is however increasingly challenging when the number of frequencies is increased, and whether this control strategy is feasible for irregular wave fields remains unclear. }

\begin{table}%
\footnotesize
\begin{tabular}{c|c|ccc}
$n$ & $\An$& $\h\eta_{10n}\oo1$  &$\h\eta_{20n}\oo1$&    $\h\eta_{30n}\oo1$       \\\hline 
1&$-0.1534 + 0.0653$i &$-0.0009 - 0.0112$i &$+0.0632 + 0.0417$i &$-0.2157 + 0.0347$i\\
2&$+0.0493 + 0.0081$i &$-0.0051 + 0.0064$i &$+0.0281 - 0.0010$i &$+0.0263 + 0.0028$i\\
3&$-0.0211 - 0.0068$i &$+0.0244 - 0.0303$i &$-0.0781 + 0.0850$i &$+0.0325 - 0.0615$i\\
4&$-0.0115 + 0.0049$i &$+0.0057 - 0.0386$i &$-0.0285 + 0.1079$i &$+0.0113 - 0.0643$i\\
5&$-0.0070 + 0.0038$i &$+0.0063 - 0.0337$i &$-0.0294 + 0.0931$i &$+0.0161 - 0.0557$i\\
\end{tabular}
\caption{Example polychromatic problem $\w_n=n\sqrt{g/h}$.
$k_{0n}|\An|=0.2$ with random phases.$\wbl_i/h=1/4$, $1/2$ and $1$ and $h=1$\,m.}
\label{tab:irr}
\end{table}

\section{Examples of fully flexible wavemakers}
\label{sec:flexi}
The arbitrary number of wavemaker hinges allows for a an increasingly flexible wavemaker, 
approximately smooth  as the number of hinges $N$ is made large.
By example, we here examine wave generation from smooth paddle profiles, distributing hinges uniformly according to $\wbl_i = i h/(N-1)$ for $i<N$ and $ \wbl_N =\infty$.
In order to match the wavemaker motion  \eqref{eq:XO1}  to a desired profile
\begin{equation}
	X(y,z,t)  = \frac12 \sum_n \xi_n(z)  \ee^{\ii (\wn t - \kyn y)} 
	\label{eq:X_xi}%
\end{equation}
we discretise the vertical coordinates $\{z_l\}$ and fit $\{\h X_{in}\oo1\}$ to
\begin{equation}
	\sum_l \h X_{in}\oo1 \frac{\max(\wbl_i+z_l,0)}{\wbl_i} = \xi_n(z_l)
	\label{eq:X_piecewise}%
\end{equation}
using linear regression  at each frequency $n$.
\\

\begin{figure}[H]
	\centering
	\subfloat[First-order solution.]{\includegraphics[height=.37\columnwidth]{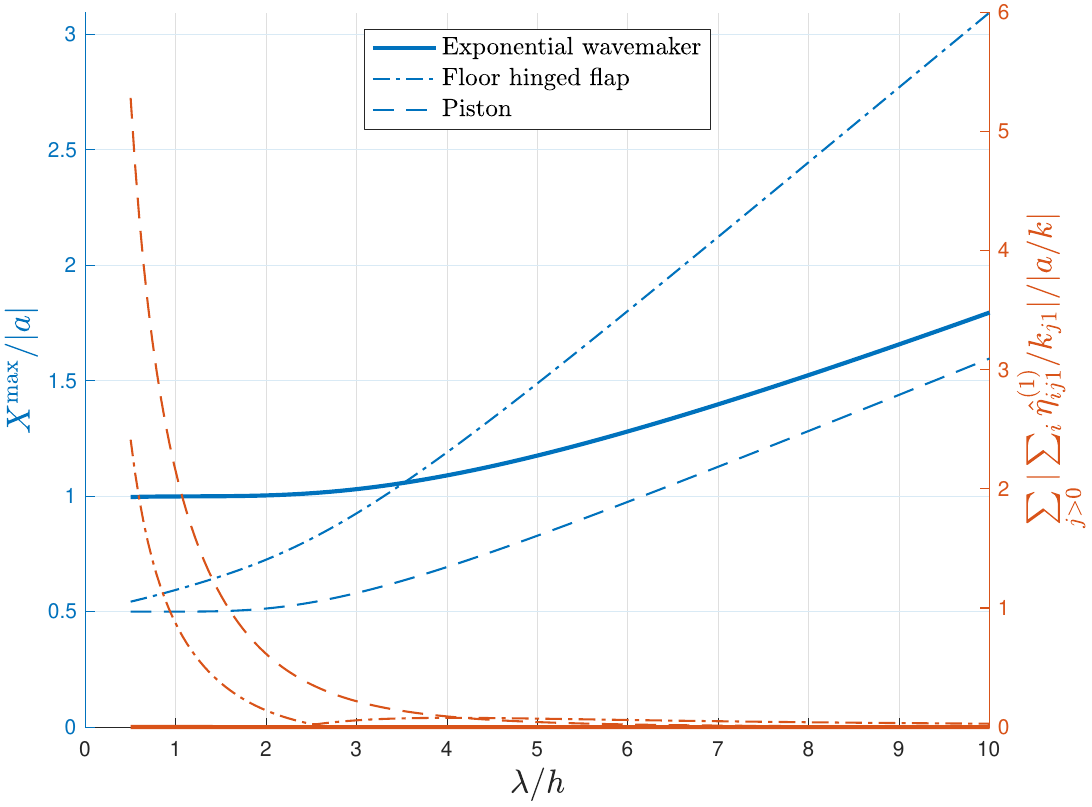}\label{fig:flex:particle:O1}}\hfill%
	\subfloat[Second-order spurious wave.]{\includegraphics[height=.37\columnwidth]{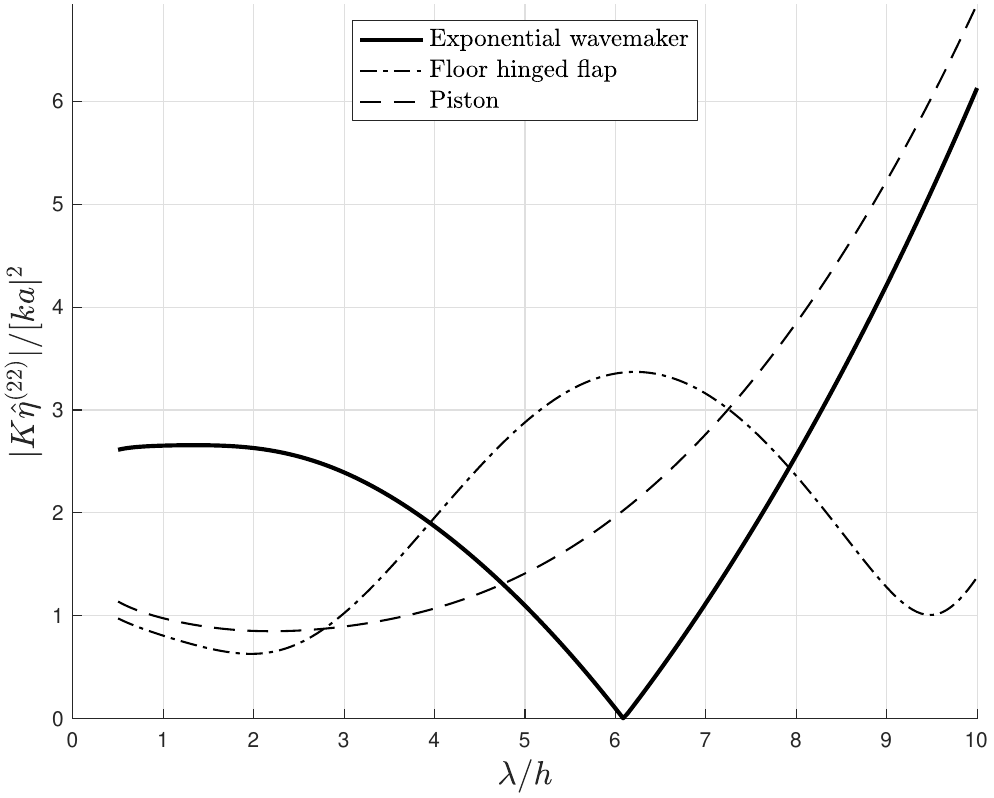}\label{fig:flex:particle:O22}}%
	\caption{Properties of the flexible wavemaker \eqref{eq:xi_particle}, which follows the linear particle trajectory. 
		Plotted in the left axes of the left panel is the peak wavemaker draft length, $X\^{max}$, relative to  wave amplitude $\A$. In the right axes of this panel is plotted the weighted measure of evanescent near-field intensity shown earlier. 
		In the right panel, the steepness of the spurious progressive wave 
		is plotted, scaled by the principal steepness squared. 
	}
	\label{fig:flex:particle}
\end{figure}

Let us here consider two examples.
First, we let the wavemaker follow the horizontal fluid particle trajectory of a regular linear wave, 
\begin{equation}
	\xi_n(z) = -\ii a \frac{\cosh k_{0n}(z+h)}{\cosh k_{0n} h},
	\label{eq:xi_particle}%
\end{equation}
thereby producing, to linear order, a wave free of evanescent modes. 
Properties of this wavemaker are shown in \autoref{fig:flex:particle}  in the limit of complete flexibility, $N\to\infty$, with piston  and flap wavemakers included for reference. 
Indeed, the magnitude measure of evanescence modes vanishes as postulated.
Peak wavemaker draft $X\^{max}$ remains comparable to the piston and flap.
Despite matching the linear wave motion,
no substantial reduction is seen in the spurious wave compared to the references.
It therefore seems that an ideal boundary must matched the particle trajectories also at second order (\eqref{eq:system_lin:Q} and \eqref{eq:defQ}).
\rev{
That being said, reducing the intensity of the evanescent near field provides benefits other than limiting spurious waves. 
A large near field will contribute to early wave breaking, thereby limiting the wave generation capacity. 
It also agitates corss-modes, setting up transverse sloshing motions that hinder wave generation further. 
Wavemakers that minimise the near field would therefore likely be of considerable  value.
}
\\

\begin{figure}[H]
	\centering
	\subfloat[First-order solution.]{\includegraphics[height=.37\columnwidth]{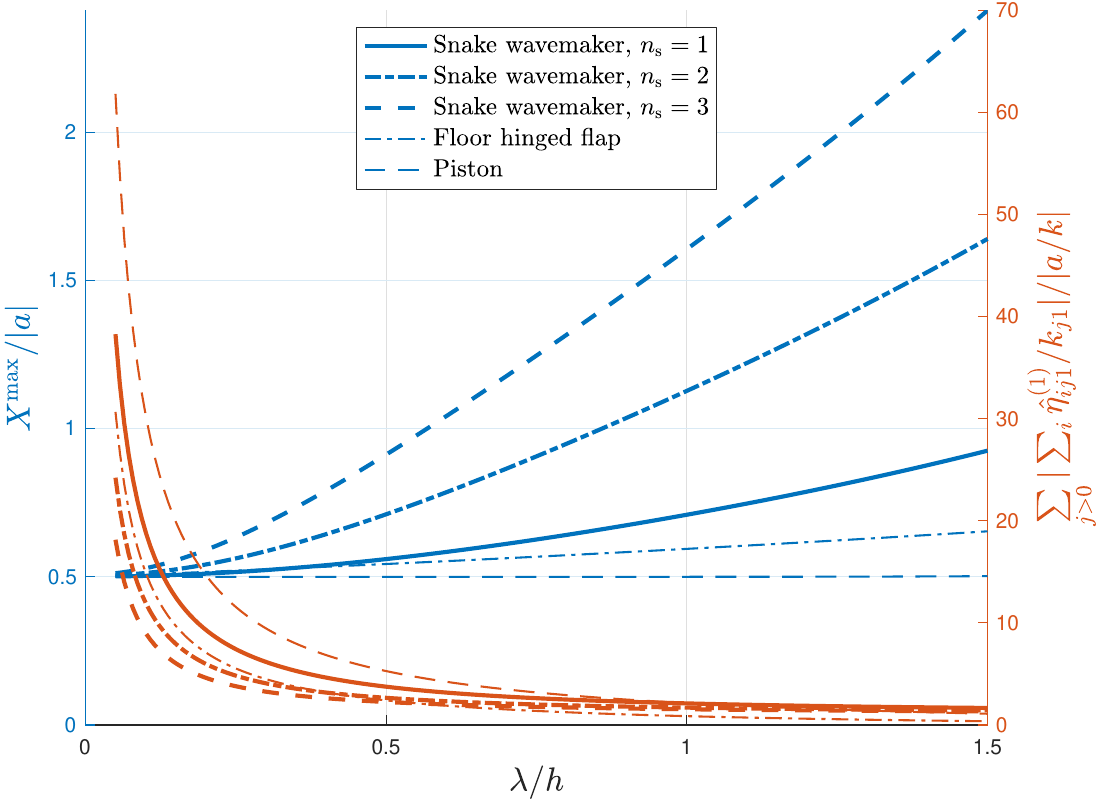}\label{fig:flex:snake:O1}}\hfill%
	\subfloat[Second-order spurious wave.]{\includegraphics[height=.37\columnwidth]{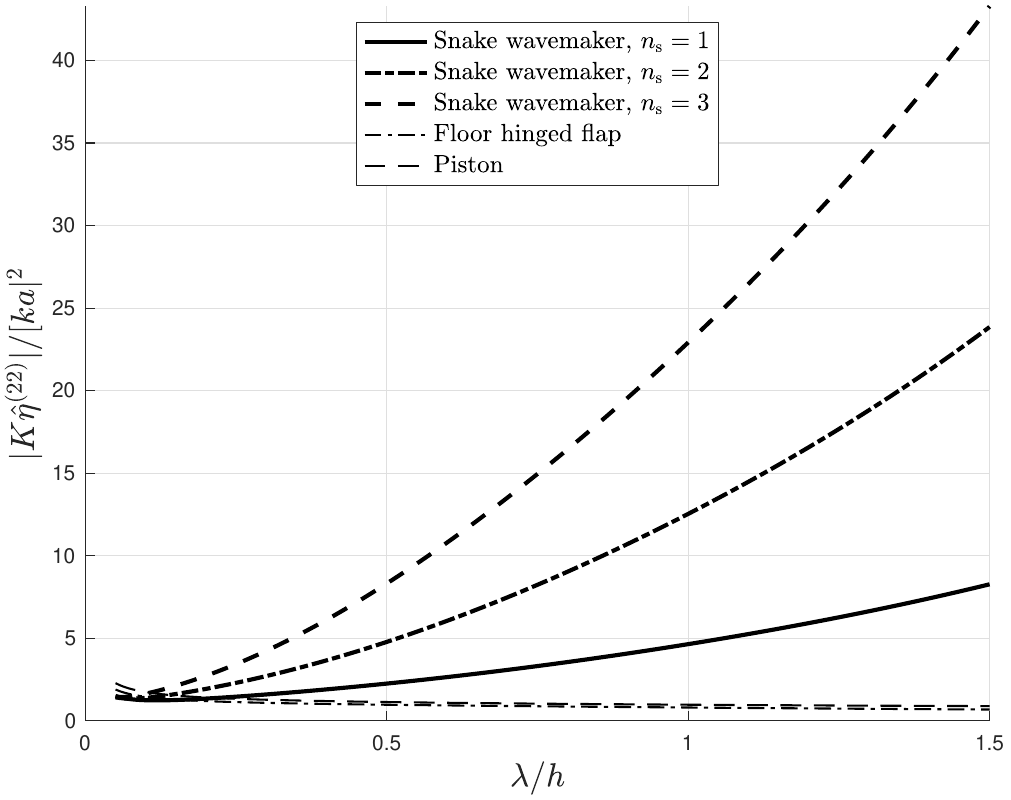}\label{fig:flex:snake:O22}}%
	\caption{As \autoref{fig:flex:particle} with wave paddle \eqref{eq:xi_snake} shaped like a snake moving vertically.}
	\label{fig:flex:snake}
\end{figure}

\begin{figure}[H]
	\centering
	\subfloat[Full potential.]{\includegraphics[height=.24\columnwidth]{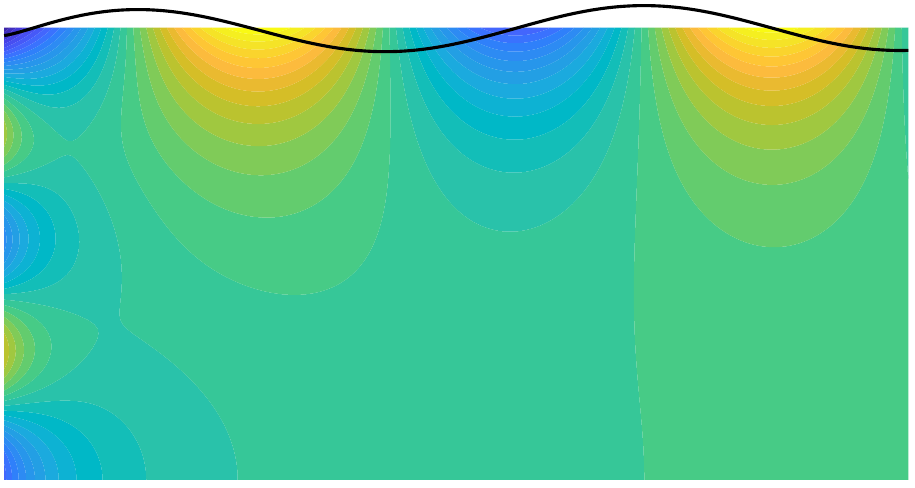}\label{fig:flex:snakeField:full}}\hfill%
	\subfloat[Potential of evanescent field.]{\includegraphics[height=.24\columnwidth]{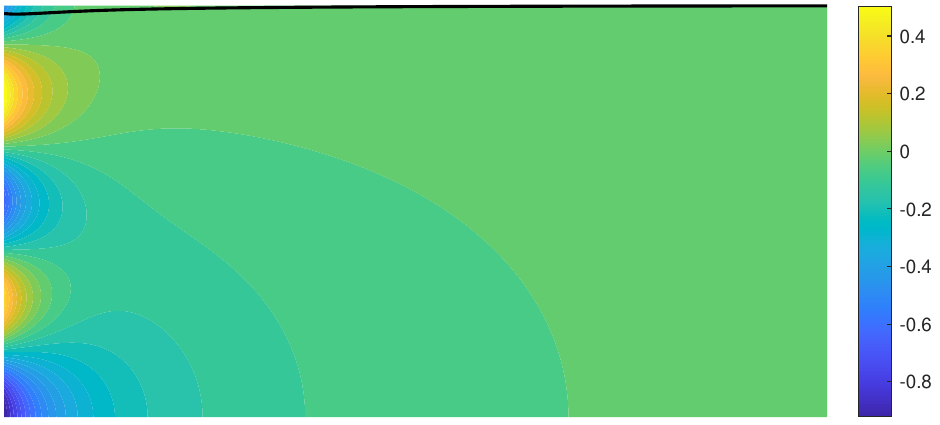}\label{fig:flex:snakeField:Ev}}%
	\caption{Example snapshot of potential field from snake-shaped paddle \eqref{eq:xi_snake}, $n\_s=2$. The evanescent field is scaled by the potential amplitude of the progressive mode. Surface elevation is plotted in black.}
	\label{fig:flex:snakeField}
\end{figure}

As a somewhat arbitrary second example, let us consider a sinusoidal wave paddle shape that snakes vertically upwards:
\begin{equation}
	\xi_n(z) = \exp(-\ii  \kappa  z ),\; \kappa= 2\pi n\_s/h.
	\label{eq:xi_snake}%
\end{equation}
Motivating the shape is the notion that something may happen as the wavelength of the paddle matches that of the progressive wave.
The resulting properties and performance is presented in \autoref{fig:flex:snake}---%
as it turns out, nothing remarkable is found and the wavemaker is seen to perform very poorly.
Extra large paddle drafts are needed and these produce spurious waves that are larger than those form the flap or the piston. 
For demonstration,  \autoref{fig:flex:snakeField} shows the spatial potential $\phi\oo1$ for when the snake wavelength is half the water depth.

\section{Summary}
\label{sec:summary}
Following \citet{schaffer2003_3D_correction}, multi-hinged, multi-directional second-order wavemaker theory has been derived and validated,
with special limits relating the  Stokes drift to the back-flow of the zeroth progressive second-order free wave. 

In previous studies, double-hinged wavemaker motions have been optimised under the conditions that both flaps move in phase. 
Under this restriction, the strategy of moving only one flap at the time (depending on the  wavelength-to-hinge-depth ratio) is favoured. 
However, flaps are allowed to move out of phase, then these 
can be optimised to completely eliminate second-order spurious waves without imposing higher harmonic motions. 
\rev{Surprisingly, the double-hinge wavemaker motion that eliminates spurious waves does not resemble a exponential profile, but rather bulges at the middle, with wavemaker flaps moving in near opposite phase.
For most periods, this opposing flap phases serves to reduce the peak wavemaker draft compared  to using only one flap.

A plausible benefit of eliminating spurious waves with the  double-paddle strategy, as compared to adding double-frequency correction, is reducing wave contamination at third order and above. 
This notion was in part motivated by \citet{fouques2022OMAE_multihinge}, who observed that  double-frequency corrections in turn contribute to third order contamination. 
Unfortunately, a brief experimental study did not reveal any reduction of third-order contamination compared to conventional correction, but remained inconclusive due measurement inaccuracy.
More precise measurements, larger wavelength-to-hinge-depth ratios or numerical studies could shed more light on the matter. 
The experiments did however successfully demonstrate single-harmonic suppression of spurious second-order free harmonics.
}

The multi-hinged wave correction strategy, demonstrated with regular waves, can in principle be extended to wave field of several frequencies or even irregular wave fields.
Three hinges or more are required in order to eliminate spurious waves provided all frequencies are uniformly spaced, zero-frequency included.
However, as the number of frequencies grows large, solving the resulting system of quadratic equations becomes challenging.
\rev{The feasibility of this correction strategy is therefore unclear concerning irregular waves.}

Increasingly flexible wavemaker  paddle profiles can be represented by increasing the number of wavemaker hinges. 
It was demonstrated that an exponential paddle profile can to linear order generate progressive waves  free of evanescent modes. 
However, both evanescent and spurious waves are found at second order, with magnitudes similar to paddles and pistons. 
\rev{A wave generation system that appears ideal in linear theory  is  therefore  unlikely to produce superior wave quality without also including nonlinearities.}
A wavemaker shaped like a vertical snake was also examined and found to perform poorly.

\section*{Acknowledgements}
This authors would like to acknowledge support for this work from Statsbygg, the Norwegian Directorate of Public Construction and Property, as part of the Norwegian Ocean Technology Centre (NHTS) project which is currently under construction in Trondheim, Norway.
The author is grateful to  S{\'e}bastien Fouques at SINTEF Ocean for support and valuable input, and to Kontorbamse for useful suggestions.

\bibliographystyle{abbrvnat}
\bibliography{sintef_bib}

\appendix

\section{Shape functions}
\label{sec:shape_functions}
\rev{
	Shape functions that appear in the wavemaker model are presented here. 
	These arise from the term
	  $\cosh k(h+z)$ in \eqref{eq:phiO1},
	 which is its own orthogonal basis enforced through the dispersion relation \eqref{eq:dispersion}.
	This orthogonality may be written
\begin{equation}
\kxjn\int_0^h \!\frac{\cosh \kjn z}{\cosh \kjn h} \frac{\cosh k_{ln} z}{\cosh k_{ln} h} \,\dd z
=\begin{cases}
	0; & l\neq j,\\
	\Lambda(\bkjn); & l=j,
	\end{cases}
	\label{eq:orthCosh}
	\end{equation}
	with kernel
	\begin{equation}
	\Lambda(\bm \kap) = \frac{\kap_{x}}{2\kap} \br{ \frac{\kap h}{\cosh^2k h} + \tanh \kap h}.
	\label{eq:Lambda}%
	\end{equation}%
}%
For ease of comparison, the notation in \citet{schaffer_1996} is adopted, introduce the lengths
\begin{equation*}
\li = \wbl_i-h
\quad\text{and}\quad
\di = \min[h,\max(0,-\li)],
\end{equation*} 
$\li$ being the downward-facing distance from hinge $i$ to bed, and
$\di$  the upward-facing distance from the bed to hinge $i$, truncated within $-h<z<0$.

The shape function 
\begin{equation*}
\frac{1}{h+\li}\int_{\di}^h \! \kap(z+\li)\frac{\cosh \kap z}{\cosh \kap h}\,\dd z
\equiv \Gamma_1(\kap)
\end{equation*}
is used for satisfying \eqref{eq:system_lin:Q} and
evaluates to
\begin{equation*}
\Gamma_{i,1}(\kap) = \sbr{\frac{z + \li}{h+\li}\frac{\sinh \kap  z}{\cosh \kap h} - \frac1{\kap(z+\li)}\frac{\cosh \kap z}{\cosh \kap h}}_{\di}^h
\end{equation*}
with $[f(z)]_{\td}^h=f(h)-f(\td)$.
The term $ \frac{\di+\li}{h+\li} \sinh\kap \di$, which appears in all the above shape functions, as well as in Sch{\"a}ffer's expressions, equals zero provided the hinge is below the still water line.
Therefore, 
\begin{equation}
\Gamma_{i,1}(\kap) = \tanh \kap  h - \frac1{\kap(h+\li)}\br{1-\frac{\cosh \kap \di}{\cosh \kap h}}.
\label{eq:Gamma1}
\end{equation}
 
The shape function 
\begin{equation*}
\frac{1}{h+\li}
\int_{\di}^h \!\sbr{  \frac{\kap_1^2-\kap_{y,1}\kap_{y,2}}{\kap_1}  (z+\li)\frac{\cosh \kap_1 z}{\cosh \kap_1 h} + \frac{\sinh \kap_1 z}{\cosh \kap_1 z}  }\frac{\cosh \kap_2 z}{\cosh \kap_2 h}\, \dd z 
\equiv \Gamma_{i,2}(\bm\kap_1,\bm\kap_2)
\end{equation*}
appearing in \eqref{eq:hphiO22} evaluates to
\begin{subequations}
\begin{align}
\Gamma_{i,2}(\bm\kap_1,\bm\kap_2) &= \frac{\sum_\pm \tilde\Gamma_{i,2}\Big(\frac{\kap_1^2-\kap_{y,1}\kap_{y,2}}{\kap_1},\kap_1 \pm \kap_2\Big)}{\sum_\pm \cosh(\kap_1 \pm \kap_2)h},\\
\tilde\Gamma_{i,2}(\kkap_1,\kkap_2) &= 
\begin{cases}
\lr[{\frac{\kkap_1}{\kap_2} \frac{z+\li}{h+\li} \sinh\kkap_2z + \br{1-\frac{\kkap_1}{\kap_2}} \frac{\cosh\kap_2 z}{\kkap_2 (h+\li)}}]_{\td}^{h};
& \kkap_2\neq0,\\ 
 \kkap_1 \frac{h-\di}{h+\li}\br{\li+\frac{h+\di}{2}}; & \kkap_2=0,
\end{cases}
\end{align}%
\label{eq:Gamma2}%
\end{subequations} 
the sum running through the pair $\pm\in\{+,-\}$.
This representation is of course equivalent to the lengthier expression presented in \citet{schaffer2003_3D_correction}
with slight variations in definition.

\section{The lateral boundary condition of multi-hinged flap wavemakers}
\label{sec:BC_x}
The lateral boundary condition for multi-hinged wavemakers is here considered in more detail.
Let the two-dimensional coordinate $\bX=(X,Z)$ describe the wavemaker as illustrated in \autoref{fig:flap}.
This is assumed  a function of the arc length $s$ running along the surface of the flaps.
Hinges at angle $\theta_i(t)$, $i = 1,2,\ldots$, are located at  positions $s_i=-\wbl_i$ at spatial coordinates $\bX(s_i,t)$.
The orientation of this arc length is chosen to coincide with the vertical $z$-axis when the flap is vertical,
i.e., $\bX(s,t) = s \be_z$ when all $\theta_i=0$. 
The flap position can then be described 
\begin{align}
\bX(s,t) &= s\be\_t(s,t) - \sum_{i=1}^N s_i\sbr{\be\_t(s_i,t)-\be\_t(s_{i-1},t)} \mc H(s-s_i),
\label{eq:bX}
\end{align}
$\mc H$ being the  Heaviside function, equalling one with a positive argument and zero otherwise,  and $s_0\to\infty$.
The tangential unit vector $\be\_t(s,t)= (\sin\vartheta, \cos\vartheta)$ depends on the cumulative hinge angle 
\begin{equation}
\vartheta(s,t) = \sum_{i=1}^N \theta_{\rm i}(t) \mc H(s-s_i).
\label{eq:vartheta}
\end{equation}
\\

Expression \eqref{eq:bX} is now simplified  to second order.
Taylor expanding $\be_t$ for small cumulative angles and limited hinge depths yields
\begin{align}
\bX(z,t) &= z \be_z+\sum_{i=1}^N   \theta_i(t) (z-s_i)  \mc H(z-s_i) \be_x + O(X \vartheta)\be_z + O(X \vartheta^2)\be_x.\label{eq:bX_lin}
\end{align}
Variable symbol $s$ has here been changed to $z$.
The expression is seen to be a  superposition of single-hinge displacements, equivalent to \eqref{eq:XO1} and \eqref{eq:XO2} with $\theta_i(t) = \frac12\sum_n \hX_i/\wbl_i \ee^{\ii \wn t}$.
Note that it is the cumulative angle $\vartheta$ and total displacement $X$ that takes the role of smallness parameters.
\\

It should be noted that expression \eqref{eq:BC:X} holds even for  extreme angles $|\vartheta|<\pi/2$, although a more general formulation of the boundary condition ($\nabla \phi \cdot \be\_n = u\_n$) is then required.

\end{document}